\documentclass[12pt]{article}
\pdfoutput=1
\usepackage{graphicx,amsmath,amsfonts,slashed}
\newcommand{\NPB}[3]{\emph{ Nucl.~Phys.} \textbf{B#1} (#2) #3}   
\newcommand{\PLB}[3]{\emph{ Phys.~Lett.} \textbf{B#1} (#2) #3}   
\newcommand{\PRD}[3]{\emph{ Phys.~Rev.} \textbf{D#1} (#2) #3}   
\newcommand{\PRL}[3]{\emph{ Phys.~Rev.~Lett.} \textbf{#1} (#2) #3}

\def\dalemb#1#2{{\vbox{\hrule height .#2pt
        \hbox{\vrule width.#2pt height#1pt \kern#1pt
                \vrule width.#2pt}
        \hrule height.#2pt}}}

 \def\bd{\begin{document}} \def\ed{\end{document}}
\def\ds{\documentstyle} \let\fr=\frac \let\bl=\bigl \let\br=\bigr
\let\Br=\Bigr \let\Bl=\Bigl 
\let\bm=\bibitem
\let\na=\nabla
\let\pa=\partial \let\ov=\overline
\def\ie{{\it i.e.\ }} 
\def\beq{\begin{equation}}
\def\eeq{\end{equation}}
\def\beqa{\begin{eqnarray}}
\def\eeqa{\end{eqnarray}}

\newcommand{\ba}{\begin{array}}
\newcommand{\tn}{\tabularnewline}
\newcommand{\ea}{\end{array}}
\newcommand{\td}{\tilde}
\newcommand{\norsl}{\normalsize\sl}
\newcommand{\ns}{\normalsize}
\newcommand{\refs}[1]{(\ref{#1})}
\def\simlt{\mathrel{\lower2.5pt\vbox{\lineskip=0pt\baselineskip=0pt
           \hbox{$<$}\hbox{$\sim$}}}}
\def\simgt{\mathrel{\lower2.5pt\vbox{\lineskip=0pt\baselineskip=0pt
           \hbox{$>$}\hbox{$\sim$}}}}

\begin{document}
\thispagestyle{empty}
\rightline{SHEP-12-37}
\vskip 1.0truecm
\centerline{\Large\bf Uncovering quasi-degenerate Kaluza-Klein Electro-Weak }
\vskip0.25cm
\centerline{\Large\bf gauge bosons with top asymmetries at the LHC}
\vskip 1.truecm
\centerline{{\large Elena Accomando}, {\large Ken Mimasu} and  {\large Stefano Moretti} }
\vskip .5truecm
\centerline{\it School of Physics \& Astronomy, University of Southampton,} 
\centerline{\it Highfield, Southampton SO17 1BJ, UK }
\vskip .5truecm

\vskip 1.truecm
\centerline{\bf\small ABSTRACT}
\vskip .5truecm
\noindent
By exploiting the correlation between charge and spin polarisation asymmetries in $t\bar{t}$, we show that combining the two observables could identify the presence of quasi-degenerate states in a resonant signal at the LHC. As an example, we investigate experimental signatures emerging in top-antitop final states in the context of a model where the Standard Model Electro-Weak sector is allowed to propagate in large extra--dimensions of TeV$^{-1}$ size while the colour sector is localised. Assuming current experimantal constraints from the 7 and 8 TeV runs and taking into account the estimated top (anti-top) reconstruction efficiencies, we find that the 14 TeV upgraded LHC with the planned integrated luminosity $L=100 fb^{-1}$ could access these quasi-degenerate multiple resonances and explore for the first time the rich phenomenology in the asymmetry observables. The main outcome would be having measurable quantities, complementary to the usual total and differential cross sections, capable of distiguishing a quasi-degenerate multiply resonant spectrum from a `standard' single resonance that could present a similar signal in a bump hunt analysis. 
\vskip 1.5truecm

\begin{flushleft}
\today
\end{flushleft}

\vfill\eject

\section{Introduction}
Extra gauge bosons are among the most common ingredients of Beyond the Standard Model (BSM) scenarios 
motivated by a variety of extensions of gauge and/or space-time symmetries. Furthermore,
resonant physics is one of the primary and straightforward sectors in which searches for such new 
physics are undertaken at modern collider experiments. It is also already well known that, aside from traditional 
differential cross section observables, more involved quantities like asymmetries can provide additional 
probes with which to analyse the properties of such objects, should they be observed at the Large Hadron Collider (LHC). 
In this paper we discuss a study of multiple neutral gauge bosons in regimes where the 
traditional `bump-hunt' searches are not sufficient to observe the presence of all of the resonant states 
due to a mass quasi-degeneracy more severe than the mass resolution of the search channel. 
Using a model of extra dimensions as an example, we find that -- once again -- asymmetries come to 
the fore and can allow for the distinction between the presence of one and  multiple (two in our case) resonant states.

The existence of large extra dimensions compactified in the TeV
range \cite{Ant}, for which the fundamental string
or quantum gravity scale is in turn rather low \cite{add}--\cite{st},
is a scenario easily testable at the LHC.
Further, if one dismisses the traditional assumption that all Standard Model (SM)
gauge bosons propagate in the same compact space \cite{IK}--\cite{Nath}
and instead allow for the more general case whereby the SM gauge structure 
arises from branes extended in different compact directions, one realises a
scenario that provides an ideal testbed for our purposes. Specifically, a general setup in which (quasi-)degenerate resonances are likely to occur is in such models of 
extra dimensions with relatively large compactification scales, $R^{-1}$.
 Allowing the gauge sector to propagate in the bulk typically results 
in strong limits on the compactification scale coming from lower mass bounds on Kaluza-Klein (KK) excitations from resonance 
searches or Electro-Weak Precision Tests (EWPTs) depending on the specific localisation of different parts of the fermion sector. 
Since the tree-level KK masses of the gauge bosons are multiples\footnote{This is true for the case of one extra 
dimension of compactification radius $R$, but depends on the specific compactification volume in the case of 
more than one extra dimension, although the compactification scales still remain the only parameters that define 
the approximate scale of the KK masses. We assume here the case of one flat extra dimension for simplicity.} of $R^{-1}$, 
one may expect that the KK EW gauge sector of such a theory would be near-degenerate since 
$R^{-1}>>g(g^{\prime})v$ where $g$($g^\prime$) and $v$ denote the $SU(2)_{L}$($U(1)_{Y}$) gauge 
couplings and the Higgs vacuum expectation value of the SM respectively. Later on, 
we will discuss the fact that particles that propagate in the bulk in such models generically incur loop-induced 
mass splittings that can be important, particularly at high compactification scales.

Within this construct, we find a realisation which complies with current stringent bounds from dijet and
$t\bar t$ events emerging after the 7 and 8 TeV runs yet remains accessible at
the 14 TeV stage. This is the one where only the EW  gauge bosons can appear
as KK excitations, but not the gluons. In addition, 
one can localise matter fermions in such a way that the production of leptonic final states
is depleted with respect to that of both light and heavy quarks, as the latter are notoriously less accessible than
the former in the LHC environment. In these conditions then, which can be realised in a Type I picture of the brane-world scenario, given that
the sensitivity of LHC data is maximal to either processes induced by Quantum Chromo-Dynamics (QCD) effects
(as opposed to those due to EW interactions) or to very clean final states involving only leptons (as opposed to both
light and heavy quarks),
one is not confronted with the very stringent bounds that would emerge if gluons (necessarily yielding
dijet and $t\bar t$ final states)
propagated in the large extra dimensions 
or EW gauge bosons propagating therein could decay in leptons. Therefore, the investigation of the effects of the 
extra-dimensional propagation of the EW gauge bosons yielding both light and heavy quarks in the final state
remains viable also in the light of the most recent data.

It is the purpose of this paper to investigate the case of the neutral EW gauge bosons, i.e., 
the $U(1)_{Y}$ and $SU(2)_{L}$ states of the SM, $\gamma$ and $Z$, and their KK excitations (or admixtures thereof), 
henceforth denoted as $\tilde{\gamma}^{\prime}$ and $\tilde{Z}^{\prime}$, respectively, produced from quark-antiquark scattering
at the LHC and yielding top-antitop pairs in the final state. After accounting for existing lower bounds on the compactification scale from direct searches in di- and $t\bar t$ data samples generated at 7 and 8 TeV, we show that
one will be able to observe at least the first excitation of the EW states at the 14 TeV stage in $t\bar t$ final states.
Further, while the extraction of information on the additional excitations would be desirable to disentangle 
the extra-dimensional model from alternative new physics scenarios, we prove that the ability of defining both charge and spin asymmetries in $t\bar t$ final
states (unlike the case of dijets) can potentially disentangle the two states (despite these appearing degenerate and unresolvable in the invariant 
mass distribution), consequently distinguishing this BSM scenario from ones involving individual resonances (like, e.g.,
$Z'$ models). Finally,  we will also illustrate that such a method can be adapted to other models showing a similar
spectrum configuration, by borrowing similar results from previous literature of ours \cite{AllOurs}.
 
The plan of the paper is as follows. In the next section we define the observables and discuss their dependence on the couplings of neutral resonances, 
which we exploit to differentiate these from single resonance models. Sect.~\ref{sec:model} describes the model that we use as an example in more detail, 
establishing a scenario that lies outside of current LHC limits. In Sect.~\ref{sec:results} we present
our findings and we conclude in Sect.~\ref{sec:summa}.
\section{Asymmetries}
\label{sec:asymmetries}
In this section we define the two asymmetry observables exploited to 
distinguish a model with degenerate resonances 
from generic scenarios containing a single resonance. The ability of 
asymmetries to go beyond simpler observables like 
differential cross sections lies in their special dependence on the 
couplings of the exchanged particles, as extensively studied in our 
previous work for the case of $s$-channel vector bosons~\cite{AllOurs}. 
In this case, overlapping resonances -- unresolvable in the positive 
definite cross section -- can induce asymmetries of different sign. As 
we will show, this feature means that the presence of multiple degenerate resonances affects the observables in a way 
which cannot be reproduced by the physics of any single resonance. Our 
study incorporates statistical uncertainties on an asymmetry observable, $A$, generically defined in terms of the number of Forward 
($F$) and Backward ($B$) events for an integrated luminosity 
$\mathcal{L}$, i.e., $N_{F}=\mathcal{L}\,\sigma_{F}$ and 
$N_{B}=\mathcal{L}\,\sigma_{B}$, as
\begin{equation}
    \delta A\equiv \delta\left(\frac{N_{F}-N_{B}}{N_{F}+N_{B}}\right)=
    \sqrt{\frac{1-A^2}{\mathcal{L}\varepsilon\sigma}}.\label{eqn:error}
\end{equation}

We also define an illustrative measure of statistical `significance' of an asymmetry 
prediction for the signal $A_{S}$ as the number of standard deviations 
it lies away from the background prediction, $A_{B}$,
\begin{equation}
	s=\frac{|A_{S}-A_{B}|}{\sqrt{\delta A^{2}_{S}+\delta
     A^{2}_{B}}}.\label{eqn:signif}
\end{equation}
within the confines of our parton-level analysis.
\subsection{Charge Asymmetry}
Charge or spatial asymmetry in collider physics is a measure of the 
symmetry of a particular process under charge conjugation. For a 
neutral current interaction, Charge-Parity ($CP$) invariance translates 
this into an asymmetry in the angular dependence of the matrix element 
for the production of a two body final state. The Tevatron, being a $p\bar{p}$ collider, is an ideal place to measure spatial asymmetries 
since the polar angle in the collider frame can more or less be 
identified with that of the Centre-of-Mass (CM) frame, modulo Parton 
Distribution Function (PDF) effects. Statistically, both incoming 
partons will be valence quarks and an absolute preferred direction can 
be unambiguously defined. The definition of a charge asymmetry at the 
LHC becomes somewhat more involved since the $pp$ initial state is 
$\it C$-invariant, necessitating the redefinition of the measured 
quantity itself. In this case, no preferred direction can be defined 
because the incoming quark will generally be a valence quark, while the 
antiquark must come from the sea. However, one can exploit the 
fact that the incoming quark will statistically carry a larger momentum 
fraction than the antiquark, resulting in a correlation between the 
boost of the $t\bar{t}$ system and the direction of the incoming quark. 
This property can be exploited in a number of ways, our choice here 
being to define the asymmetry with respect to the angle 
$\theta^{\ast}$: the angle in the CM frame between the outgoing lepton 
and the $z$-axis defined, on an event by event basis, to be the 
direction in which the $t\bar{t}$ system is boosted~\cite{Krohn:2011tw}. This quantity, which we call $A^{\ast}_{FB}$, is thus defined as follows:
 \begin{align}
 A^{\ast}_{FB}=\frac{N_{t(\bar{t})}(\cos\theta^{\ast}>0)-N_{t(\bar{t})}
     (\cos\theta^{\ast}<0)}{N_{\rm Total}},\label{eqn:asy_AFB}
 \end{align} 
where $N_{t(\bar{t})}$ denotes the number of tops(antitops) observed in the forward ($\cos\theta^\ast>0$) or backward ($\cos\theta^\ast<0$) direction 
and ${N_{\rm Total}}$ is the total number of events. In QCD, the 
asymmetry for the $t\bar{t}$ final state is generated dominantly at 
Next-to-Leading Order (NLO) via interference of leading order 
$q\bar{q}\to t\bar{t}$ with the corresponding box diagram as well as by the interference between initial and final state gluon radiation~\cite{QCDasymmetry}. There are also genuine tree-level EW 
contributions as well as mixed EW and QCD effects at NLO \cite{tt-EW}.

\subsection{Spin Polarisation}
One of the benefits of the $t\bar{t}$ final state is the fact that, as particles that decay before hadronising, several observables can be defined 
that probe the helicity structure of one or both of the outgoing (anti)tops. The most powerful such observable is the spin polarisation, $A_L$, 
or single spin asymmetry, defined as follows:
\begin{equation}
 A_{L}=\frac{N(-,-) + N(-,+) - N(+,+) - N(+,-)}{N_{\rm Total}},\label{eqn:asy_AL}
\end{equation}
where  $N$ denotes the number of observed events and its first(second) argument corresponds to the helicity of the final state particle(antiparticle). 
It singles out one final state particle, comparing the number of its positive and negative helicities, while summing over the 
helicities of the other antiparticle (or vice versa). The observable it traditionally extracted as a coefficient in the angular distributions of the decay products of the parent top (anti)quark~\cite{tt-pol}.

\subsection{Reconstruction}\label{subsec:reco}
While the $t\bar{t}$ channel offers a wide choice of observables that 
are sensitive to new physics, one of the primary complications of such 
analyses is the difficulty in reconstructing the 6-body final state 
that results from the pair production of tops. Ideally, one would 
perform a full chain of event generation, showering and hadronisation, 
culminating in a detector simulation to get an accurate representation 
of the reconstruction process for observables of interest. The 
associated efficiencies will depend on the information required for the 
observable and the particular decay channel of the $t\bar{t}$ system. Since our analysis is limited to be at parton level, without subsequent 
decay of the tops, it is necessary for us to employ reasonable 
estimates of reconstruction efficiencies such that our qualitative 
predictions correspond better to the reality of a detector environment. 
We estimate this quantity in a conservative manner by gauging the 
efficiencies of the primary requirements of each observable in each 
decay channel and using a net efficiency weighted by the branching 
fractions.

The common experimental requirement between the two asymmetry 
observables of interest and also the invariant mass distribution is a 
full reconstruction of the $t\bar{t}$ system. The only extra information needed for the asymmetries is the angular distributions of the decay products of one or two the tops when extracting the top spin observables. An important consideration for the analysis of new physics at several TeV is the 
likely boosted nature of the final states which will have an impact on 
the reconstruction process. The collimation of decay products means 
that many traditionally reliable measurements such as $b$-tagging, 
invariant mass reconstruction and isolation become hampered and must be 
adjusted. A variety of pruning and jet substructure methods are applied 
at the LHC~\cite{boosted} and quote efficiencies of about 30-40\% to 
tag a hadronic top and a number of analyses have used such methods in 
recent resonance searches~\cite{boosted-resonance}, showing that 
including the boosted methods increases sensitivity to higher $Z^{\prime}$ 
masses. The weighted efficiencies are quoted to be around 5 or 6\% from 
each of the fully hadronic and semi-leptonic channels. As yet, we are 
not aware of any asymmetry measurements nor analyses in the dilepton 
channel using these techniques. We therefore choose a total 10\% 
efficiency as a conservative estimate to reconstruct high mass 
$t\bar{t}$ events.

The charge asymmetry measurement can be made in any of the three 
$t\bar{t}$ decay channels and a reconstruction of the top four momenta, 
after potential top-tagging using boosted methods, is sufficient to 
obtain the quantity and nothing extra is needed beyond sufficient 
statistics to represent it as a function of $M_{t\bar{t}}$. We 
therefore use the same reconstruction efficiency estimate for this 
observable as used in the resonance searches. The top polarisation 
asymmetry is more complicated due to the need for reconstructing the 
angular distributions of decay products. What is clear is that the 
boosted systems will inhibit the measurement of such a quantity as the 
collimation of the decay products approaches the angular resolution of 
the calorimeters. At this stage, a lack of experimental analyses makes 
it difficult to estimate how well such a quantity can be measured at 
high $p_{T}$ although a number of papers discuss the problem and pose 
potential solutions moving away from the requirement of fully 
reconstructing the decay products~\cite{boosted-pol}. For this study, 
we reduce the $A_{L}$ efficiency estimate to 5\%, in lieu of a complete 
analysis which we feel is beyond the scope of this paper. As with the 
other observables, we present the spin polarisation binned in invariant 
mass to display certain features although we do not claim that this 
will definitely be possible at the LHC. However, we feel this will 
not greatly affect the conclusions of this study since the capacity 
to distinguish degenerate resonances relies mainly on integrated 
rather than differential asymmetry measurements.

\subsection{Asymmetries and resonance couplings}\label{subsec:asycoup}
Here, we elaborate on the specific coupling dependence of the asymmetries as discussed in~\cite{AllOurs} and the expectation for multiple resonances. 
The unique coupling structure of the asymmetries can be traced to the fact that they access a parity asymmetric combination of left and right-handed 
$\tilde{\gamma}^{\prime} , \tilde{Z}^{\prime}$ couplings, $C_{R}^{2}-C_{L}^{2}$, as opposed to a cross section $\sigma$, which depends only on the symmetric combination, $C_{R}^{2}+C_{L}^{2}$. 
For a given initial state with chiral couplings $q^{i}_{R,L}$ to the $\tilde{\gamma}^{\prime}, \tilde{Z}^{\prime}$, the dependence of the observables is summarised for 
the $t\bar{t}$ final state as:
\begin{align}
	\label{eqn:couplings}
	\begin{split}
	\sigma&\propto\left((q^{i}_{R})^{2}+(q^{i}_{L})^{2}\right)\left(t_{R}^{2}+t_{L}^{2}\right),\\
	A_{FB}&\propto\left((q^{i}_{R})^{2}-(q^{i}_{L})^{2}\right)\left(t_{R}^{2}-t_{L}^{2}\right),\\
	A_{L}&\propto\left((q^{i}_{R})^{2}+(q^{i}_{L})^{2}\right)\left(t_{R}^{2}-t_{L}^{2}\right).
	\end{split}
\end{align}
Naturally, the fact that the cross section is positive definite while the two asymmetries are not (as intimated already), being additionally sensitive to the relative 
`handedness' of the couplings, suggests that multiple resonances will be able to produce unique effects 
that cannot be reproduced by any single resonance. Furthermore, interference effects of the form
\begin{equation}
	\label{eqn:coup_int}
	\propto\left(q^{(1)}_{R}q^{(2)}_{R}\pm q^{(1)}_{L}q^{(2)}_{L}\right)\left(t_{R}^{(1)}t_{R}^{(2)}\pm t^{(1)}_{L}t^{(2)}_{L}\right),
\end{equation}
depending on the observable, can have a non-trivial structure, as induced by the specific couplings of the virtual objects. 
In essence, the effects of having two particles with different couplings and hence different widths can  induce interesting 
lineshape effects in the asymmetry observables while still approximating a Breit-Wigner shape in the differential cross section.

\section{The model}
\label{sec:model}
A large amount of theoretical and phenomenological literature exists on models which place the whole SM particle 
content~\cite{ued} or sometimes only its gauge sector~\cite{Ant2,scc} in the bulk. The main difference between the two being the delocalisation
of fermions which requires an orbifold compactification in order to obtain chiral states. These can be seen as extensions of the 
Arkani-Hamed--Dimopoulos--Dvali (ADD) scenario, 
which reformulates the hierarchy problem by allowing gravity to live in the bulk while localising the rest of the SM on a brane. 
The framework for a model where a selection of the SM gauge structure is allowed to propagate in the bulk is motivated in~\cite{Elena} 
and represents a mixture of the two pictures. Given the choice of localising any combination of the gauge groups and matter 
representations, a number of combinations are possible. Our study lends itself to the $(t,l,l)$ realisation 
of~\cite{Elena} (henceforth AADD), where $t,l$ denote `transverse' and `longitudinal' and refer to the orientation of the
($SU(3)_{C}$, $SU(2)_{L}$, $U(1)_{Y}$) gauge groups with respect to the extra dimension. This implies that the colour sector 
is localised while the EW one propagates in the bulk, gaining KK excitations. In order to realise a model with scales accessible at the LHC, 
the leptonic sector is also allowed to propagate in the bulk. The orbifold compactification necessary to accommodate fermions in the bulk 
preserves KK-parity, suppressing the interactions of the EW KK resonances with the leptonic sector. This simultaneously removes 
the traditional di-lepton channel from searches for such resonances and limits the constraints 
from EWPTs that typically arise from a fully localised fermion sector. In addition, having kept the quark sector 
localised along with the gluons leads to an enhancement of the couplings of the KK resonances to quarks 
relative to its SM zero-modes as a result of the KK expansion procedure. Ultimately, we are left with a model in which EW gauge bosons 
have KK excitations, $\tilde{\gamma}^{\prime}$ and $\tilde{Z}^{\prime}$, which couple universally to the quark sector with an enhancement of $\sqrt{2}$ to their SM gauge quantum numbers 
and have loop-suppressed interactions with the lepton sector which we neglect. As far as their interactions with quarks are concerned, these particles 
are heavy copies of their SM counterparts. We assume that EW Symmetry Breaking (EWSB) takes place in the bulk but that these contributions are small 
compared to the compactification radius as discussed in the introduction and we elaborate on the assumption of quasi-degeneracy in 
the next section. We therefore compute the tree-level widths of the resonances assuming only contributions from quarks with a small ($\sim 3\%$) k-factor to account for NLO QCD contributions.

We wish to use this specific realisation of an extra-dimensional model, compatible with current LHC limits, 
as an example of the scenario in which asymmetries can be used to deduce the presence of quasi-degenerate 
resonances beyond the mass resolution of the search channel. In this case, although the dijet channel represents
a more sensitive mode with respect to the signal as shown in Sect.~\ref{subsec:LHC_constraints}, we would like to
consider $t\bar{t}$ due to the fact that one can measure both its charge and polarisation asymmetries, which turns out to be essential 
in identifying the presence of more than one  particle. In any case, one would not expect the mass resolutions 
of both channels to differ greatly at such high $p_{T}$ and, further, the large uncertainties associated with jet energy scale 
are likely to further compromise the ability to resolve nearby peaks in both invariant mass spectra.

\subsection{Radiative mass corrections and mixing}\label{subsec:mass_corrections}
A typical feature of `universal' type models of extra dimensions, where some of the SM matter 
content is allowed to exists in the bulk, is that KK excitations receive radiative mass corrections 
beyond those that occur in a 4-Dimensional (4D) realisation. Considering one extra dimension for simplicity, these corrections originate from the 
violation of 5-Dimensional
(5D) spacetime symmetries caused by the compactification of the extra direction~\cite{Cheng:2002iz}.
5D loop contributions which do not break these symmetries will simply contribute to the field strength renormalisation of the 5D fields. 
Specifically, a circle compactification violates Lorentz invariance at long distances and can accommodate loop contributions with non-zero winding number 
around the extra-dimensional space and yield universal, finite corrections to the two point function proportional to $\frac{1}{R^2}$ and independent of KK number. 
Furthermore, the orbifold projection induces yet more contributions arising from the orbifold 
fixed points which violate translational invariance. Therefore, loop diagrams where a particle encounters such a 
boundary and flips its 5D momentum will also induce logarithmic corrections proportional to the KK mass $\frac{n}{R}$. The two types of corrections are termed `bulk' and `orbifold' respectively and contribute only to the 5th component 
of the field strength renormalisation factor which, upon KK decomposition of the action, corresponds to a mass correction to the 4D KK modes.

Consequently, the assumption that the gauge boson excitations at each KK level will essentially 
be degenerate with a mass of $\frac{n}{R}$ is not necessarily a good one, depending on the particular 
realisation of the model. The indirect importance of such mass splittings lies in the subsequent modification 
of the mixing between the neutral gauge bosons $\tilde{\gamma}^{\prime}$ and $\tilde{Z}^{\prime}$
which will, in turn, affect the exact coupling structure of the mass eigenstates. 
While at LO one can assume that the mixing between the hypercharge and $T_{3}$ gauge bosons 
will proceed identically to the SM with EWSB ($\theta=\theta_{W}$, where $\theta$ is the mass mixing 
angle between the resonances in AADD and $\theta_{W}$ is the Weinberg angle), mass splittings will drive the mixing back towards 
the pure gauge states and invalidate the assumption that such resonances will couple like `copies' 
of the SM $\gamma$ and $Z$ stated in~\cite{Elena}. That said, in our case, the gauge bosons 
of interest do not interact strongly, which ensures that the splitting effects will not be too large. 

For the `Universal Extra Dimensions' (UED) 
realisation\footnote{A model where the full SM particle spectrum is allowed to propagate in the bulk~\cite{ued}.} addressed in~\cite{Cheng:2002iz}, 
the aforementioned corrections to the neutral gauge sector masses 
result in a mass splitting of about 6\% of the compactification scale, 
$R$. The case of AADD closely resembles a universal scenario with 
regards to the EW sector, the only 
difference being that the localisation of quarks makes them couple 
universally to all KK modes. Thus the mass corrections to each KK level 
will resemble those of UED with the 5D quark contribution removed and 
replaced by a normal 4D SM vacuum polarisation with enhanced couplings. 
As shown in~\cite{Cheng:2002iz}, fermions do not contribute to the 
gauge boson masses via orbifold corrections which are dominant over the 
bulk corrections for all KK-levels, particularly with increasing 
$R^{-1}$ meaning that localising quarks does not have a big effect on 
the mass splitting. One would also expect a negative logarithmic 
contribution from the localised fermion interaction of each gauge boson 
proportional to $g^{\prime 2}\sum_{q}Y_{q}$ and $g^2\sum_{q}T(f)$ 
respectively, where $Y$ denote hypercharge and $T(f)$ denotes the trace 
of the generators $Tr[t_{A}t_{B}]$ in the fundamental represenation of 
$SU(2)$. We have calculated that the corrections are small compared to 
those arising from the bulk particle content and decrease 
the mass splitting by about 1\%. It is fair to say that this keeps the model within the quasi-degenerate regime since we don't expect the mass resolutions of the $t\bar{t}$ or dijet channels to be much better than 5\%. The splitting are, however, large enough to significantly affect the mixing structure of the KK EW gauge boson couplings.

Ultimately, in the context of using asymmetries to probe observed resonances in the $t\bar{t}$ spectrum, it is 
evident that having too large mass splittings will first and foremost reduce the problem to a study of multiple
single resonances as opposed to a quasi-degenerate spectrum. We would therefore like to consider the 
regime where the mass splitting could be large enough to induce non SM-like mixings (and therefore couplings) 
while maintaining a quasi-degeneracy in the first KK level so that the $t\bar{t}$ mass resolution does not permit one to fully 
resolve the two resonances in the cross section. This is chiefly because we would like to highlight the efficacy of 
using differential asymmetry observables to distinguish such a case from a single resonance in a way that is 
not possible using a differential cross section analysis. In models with a large enough mass splitting, regular 
resonance search methods will be sufficient to recognise the presence of two new bosons while, if not, an analysis 
of asymmetries will do so. We choose to present a number of results for 
the illustrative limit of fully degenerate resonances as a `worst case 
scenario' for our purposes while also including some observables for 
the spectrum with radiative corrections. 

An important point to make is that, while mass splittings will affect 
the mixing of the KK resonances, in the exactly degenerate limit, the 
mixing angle, $\theta$, should not be a physical observable around the 
resonance peak. This is clear since the mixing of two degenerate states 
simply amounts to a redistribution of couplings which can only yield 
differences in widths coming from (small) top mass effects. With this 
principle in mind, we found that it was extremely important to include 
off-diagonal widths in order to prevent artificial effects
arising when varying mixing angles. When multiple 
resonances have common decay channels and a mass splitting comparable 
to their intrinsic decay widths, it may occur 
that imaginary parts of one-loop diagrams mixes the two states via 
their width~\cite{Giacomo}. In this case, the propagators 
must be treated as a matrix with the off diagonal components from these 
loops potentially altering their resonant structure. The size of these 
effects is maximised in the degenerate limit and we find that including 
these effectively removes the mixing angle as a physical parameter up 
to (small) interference effects with the SM and higher KK gauge bosons.
In order to highlight these points, we simulate the phenomenology of 
the neutral KK resonances in both extreme cases: SM like couplings 
$\gamma^{\prime}$ and $Z^{\prime}$ ($\theta=\theta_{W}$) and maximally 
`unmixed' gauge states $W_{3}^{\prime}$ and $B^{\prime}$ ($\theta=$0), 
which turn out to show large differences in the asymmetry observables 
when not including the off diagonal effects. Since the unmixed limit 
corresponds in a sense to the restoration of the EW gauge symmetry, one
would expect the off diagonal effects to vanish in this limit. As such, 
the phenomenology of the unmixed case corresponds to the `true' 
observable while artifacts from not including off diagonal effects 
will arise once the mixing angle is switched on.

\subsection{LHC limits on $R^{-1}$}\label{subsec:LHC_constraints}
The nature of the model ensures that the new resonances couple in an enhanced manner to quarks 
while simultaneously having suppressed couplings to leptons. This dictates that the strongest 
constraints on the model will not come from EWPTs nor traditional di-lepton resonance searches 
but rather from dijet and possibly top-antitop searches. With this in mind we would like to estimate 
the current limits on the compactification scale, $R^{-1}$, using the most recent LHC (CMS) analyses 
available in the two channels, in order to use a reasonable value for this parameter in our study.
We use the latest dijet resonance search for $\sqrt{s}=$8 TeV and 19.6 fb$^{-1}$~\cite{CMS:2012eza} while for $t\bar{t}$
we found the most constraining analysis to be the boosted resonance search in the lepton+jets 
channel at $\sqrt{s}=$7 TeV~ with full luminosity~\cite{CMS:2012xva}. 

Such searches determine limits on the enhancement of the `unfolded' $t\bar{t}$ production cross section 
in the case of the lepton+jet search and $\sigma\times BR(Z^{\prime}\rightarrow j\bar{j})\times\mathcal{A}~(\text{Acceptance})$ 
for the dijet search. Both use a `bump-hunt' binned analysis fitting the background plus a single-resonance signal 
shape with the cross section as a free parameter. Consequently, the analysis is rather sensitive to the signal shape. 
The fact that any interference effects are \emph{a priori} neglected in model independent limits means that the 
limits we can obtain on our model will be in the approximate case of degenerate resonances not interfering with the SM gauge bosons, in order to best match the assumed signal shape. 
We therefore compute the production rate in our model as a function of $R^{-1}$ which we equate with 
$M_{\tilde{\gamma}^{\prime}}\approx M_{\tilde{Z}^\prime}$ and compare these predictions with the CMS data to obtain a qualitative, 
yet instructive, limit on the compactification scale. In addition to neglecting the interference effects, 
which are indeed small compared to the QCD background, we also only consider the first KK level of resonances when computing the signal cross sections.
This is also to best match the model signal shape used in the experimental analyses. The effects of the higher KK resonances are 
strongly reduced at high scales ($\geq$2 TeV) due to low parton luminosities while at the lowest scales ($\sim$1 TeV) the 
first resonance is enough to exclude the model. We note that, within these simplifications, the production rates between the SM-like mixed
and unmixed cases do not differ significantly even without including the aforementioned off-diagonal width effects. For the dijet analysis, an important additional contribution will arise from KK W-boson contributions as well as $t$-channel exchanges of all possible new gauge bosons. The former will contribute to the signal cross-section while we argue that the latter will be present as a continuum correction and would thus be absorbed into the normalisation of the background fit. As such, we only consider $s$-channel exchanges of KK gauge bosons to contribute to the visible signal cross section. Furthermore, an additional kinematical cut of pseudorapidity separation between the jets $\Delta \eta_{jj}<$1.3 is imposed along with the requirement that both jets be central ($|\eta|<$2.5). 

\begin{figure}[!h]
\centering
\includegraphics[angle=0,width=0.45\textwidth]{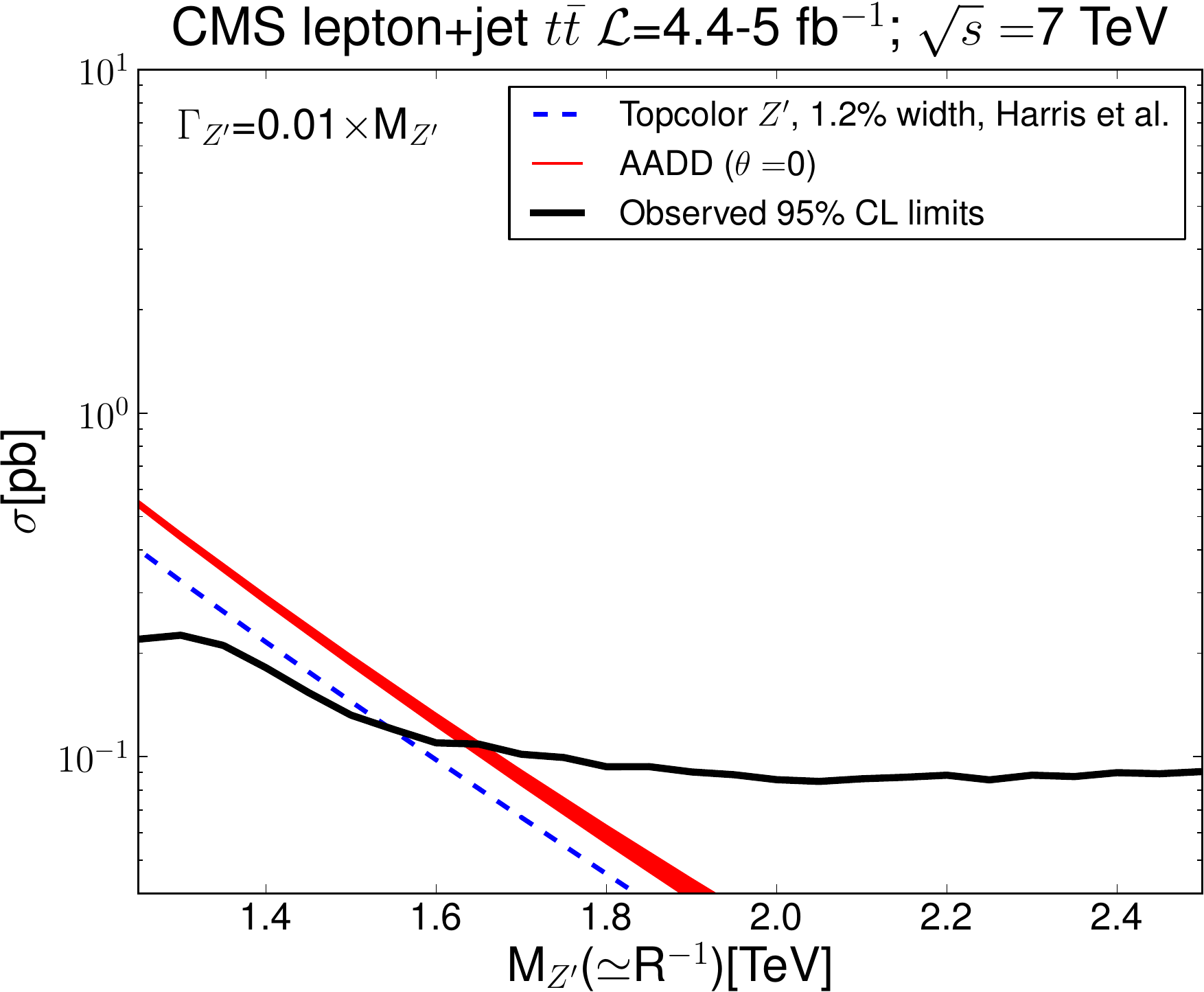}
\includegraphics[angle=0,width=0.45\textwidth]{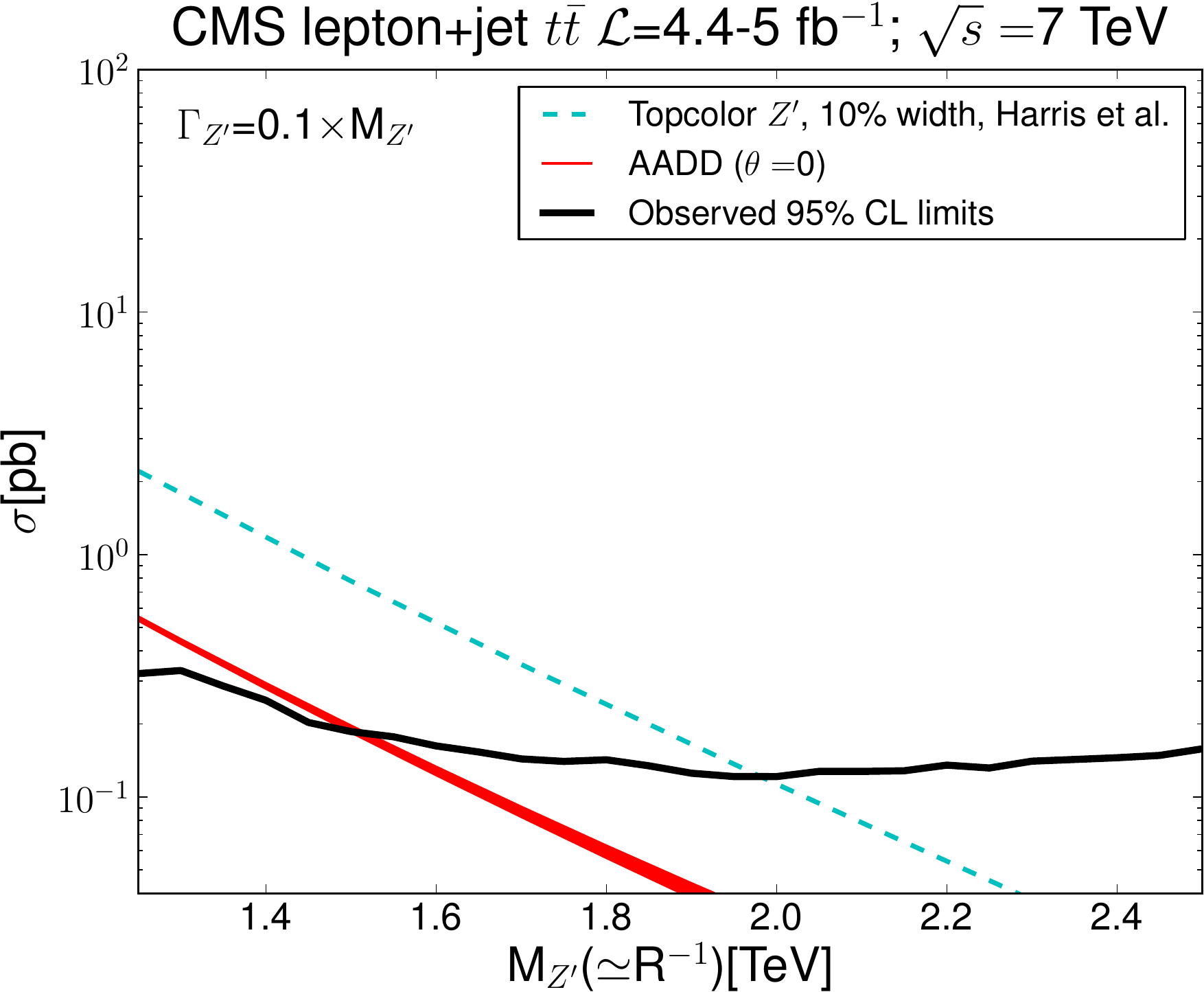}\\
\vspace{0.5cm}
\includegraphics[angle=0,width=0.45\textwidth]{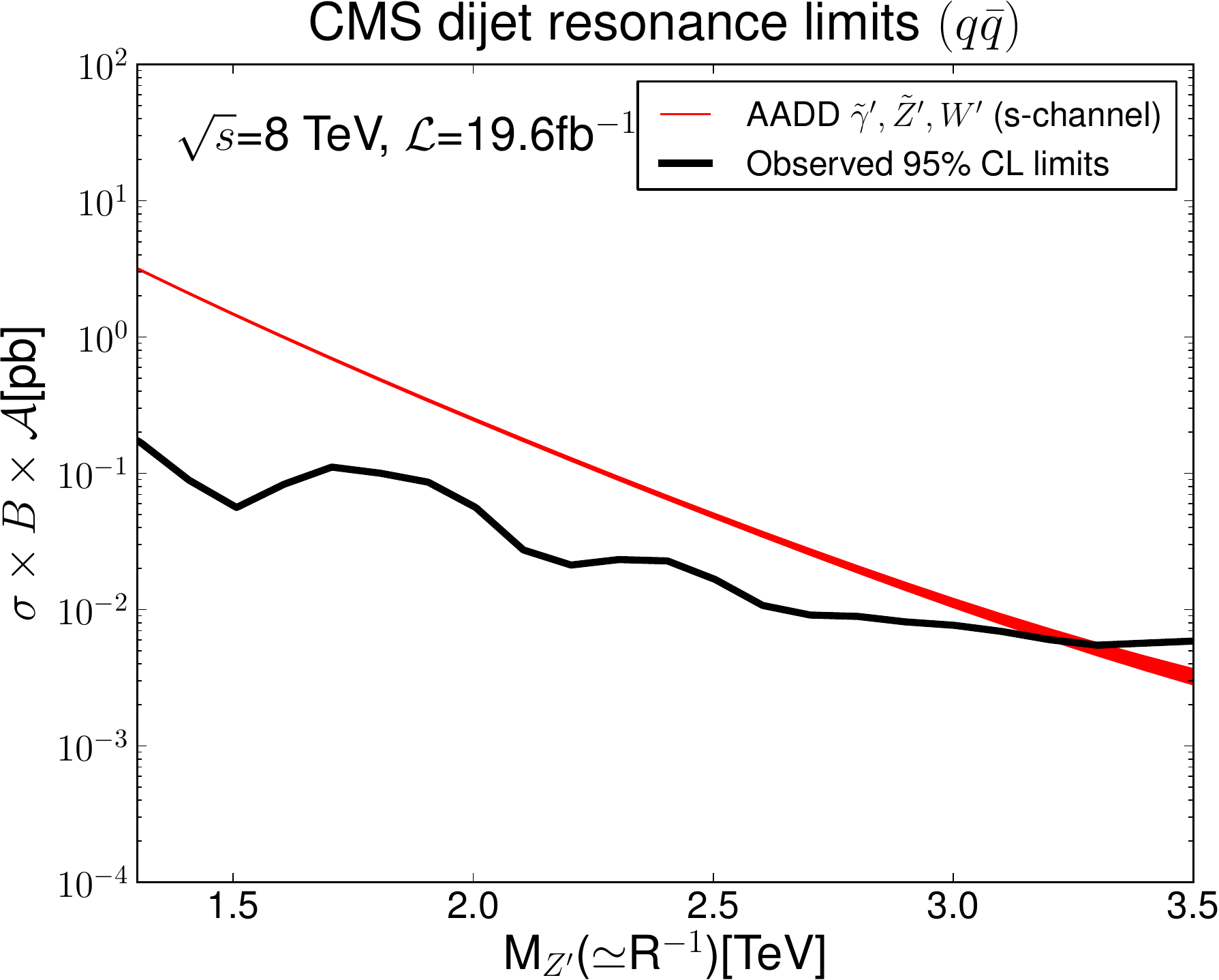} 
\begin{minipage}[b]{0.45\textwidth}
	      \hspace{0.1\linewidth}
      \begin{minipage}[b]{0.9\linewidth}     
 \caption{\label{fig:exclusions}CMS exclusion plots from the $t\bar{t}$ ({upper}) and  dijet ({lower}) resonance 
searches at $\sqrt{s}=$7 and 8 TeV, respectively. The $t\bar{t}$ exclusions assume either narrow or wide ($\Gamma_{\tilde{\gamma}^{\prime}(\tilde{Z}^\prime)} =$ 0.01 and 
0.1 $\times\,M_{\tilde{\gamma}^{\prime}}(M_{\tilde{Z}^\prime})$, respectively) scenarios compared to the Topcolor benchmark. AADD signal rates include 
statistical uncertainties.}
\vfill
     \end{minipage}
 \end{minipage}
\end{figure}

In Figure~\ref{fig:exclusions}, we compare the $t\bar{t}$ and dijet production rate in AADD to the limits 
quoted from CMS resonance searches in the two channels. The dijet rates are unsurprisingly large since the resonance 
couples with a factor $\sqrt{2}$ larger than the SM case leading to a limit of order 3.1 TeV on $R^{-1}$. 
The fact that this analysis was performed on 8 TeV data compared to 7 for $t\bar{t}$ along with the higher 
multiplicity of light quark final states and better reconstruction efficiency suggests that the latter analysis 
will not be able to compete in setting such limits. The $t\bar{t}$ limits are based on particular 
assumed widths (1\% and 10\% of the mass) of the resonances. The popular `Topcolor'~\cite{topcolor} benchmark model 
that is constrained in this analysis has been left on the figures for comparison. Given that, in our scenario, 
the tree-level width contributions come only from quarks and give a contribution of about 5\% of the mass, we compare the predictions to both cases, 
understanding that the true limit will lie somewhere in between. It appears that the exclusion 
is rather sensitive to this assumption since, in the narrow case, AADD rates are higher than the Topcolor 
ones while in the wide case they are lower, which may well be a direct consequence of the $\sim$5\% 
widths. This channel produces a limit on $R^{-1}$ of about 1.5-1.7 TeV, which is much lower than the dijet case at 8 TeV, as expected. 
We therefore choose to simulate subsequent results for a 
compactification scale of 3 TeV in order to present the phenomenology 
of the AADD model.

\section{Results}
\label{sec:results}
We now present our numerical results for the phenomenology of the AADD model as our benchmark for a quasi-degenerate 
two-resonance scenario preferentially coupled to $t\bar{t}$. As suggested by Subsects.~\ref{subsec:mass_corrections} and~\ref{subsec:LHC_constraints}, 
a compactification scale of $R^{-1}=$ 3 TeV is chosen as our reference point. 
The code exploited for our study is based on helicity amplitudes, defined through the HELAS subroutines~\cite{HELAS}, 
and built up by means of MadGraph~\cite{MadGraph}. Initial state quarks have been taken as massless whereas for the final 
state top (anti)quarks we have taken $m_t=$ 175 GeV. The CTEQ6L1~\cite{cteq} 
PDFs were used with factorisation/renormalisation scale set to the compactification scale, $Q=\mu=R^{-1}$. 
VEGAS~\cite{VEGAS} was used for the multi-dimensional numerical integrations. In each case, the BSM signal including (small) inteference with the EW zero modes ($\gamma$,$Z$) is laid against the tree level SM 
background dominated by QCD and supplemented by EW production for completeness, all at LO. We focus on differential cross section and asymmetry observables binned around the resonance peak region in invariant mass, $|M_{t\bar{t}}-R^{-1}|<$ 500 GeV. The results should not, qualitatively, be affected by 
the choice of $R^{-1}$. We will begin by showing results for the exactly degenerate limit and highlight the importance of including off-diagonal effects before moving onto the radiatively split spectrum. We will then present a comparison of the degenerate AADD model with generic single $Z^\prime$s in the asymmetry observables to underline the fact that they can be very useful in identifying the presence of quasi degenerate, multiple resonances when these cannot be resolved in the invariant mass spectrum.

\subsection{Invariant mass and asymmetry spectra}
\begin{figure}[h!]
	\centering	
	\includegraphics[width=0.4\linewidth]{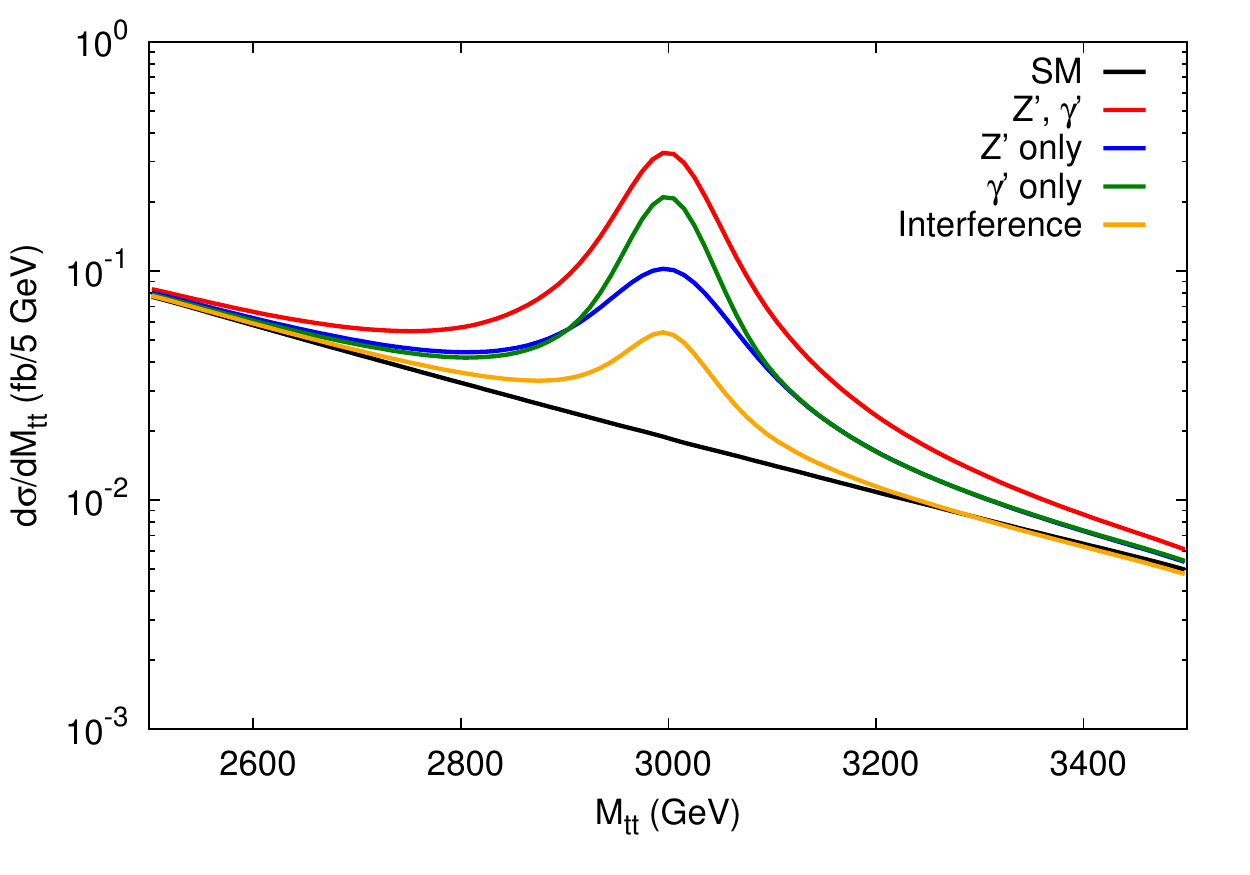}
	\includegraphics[width=0.4\linewidth]{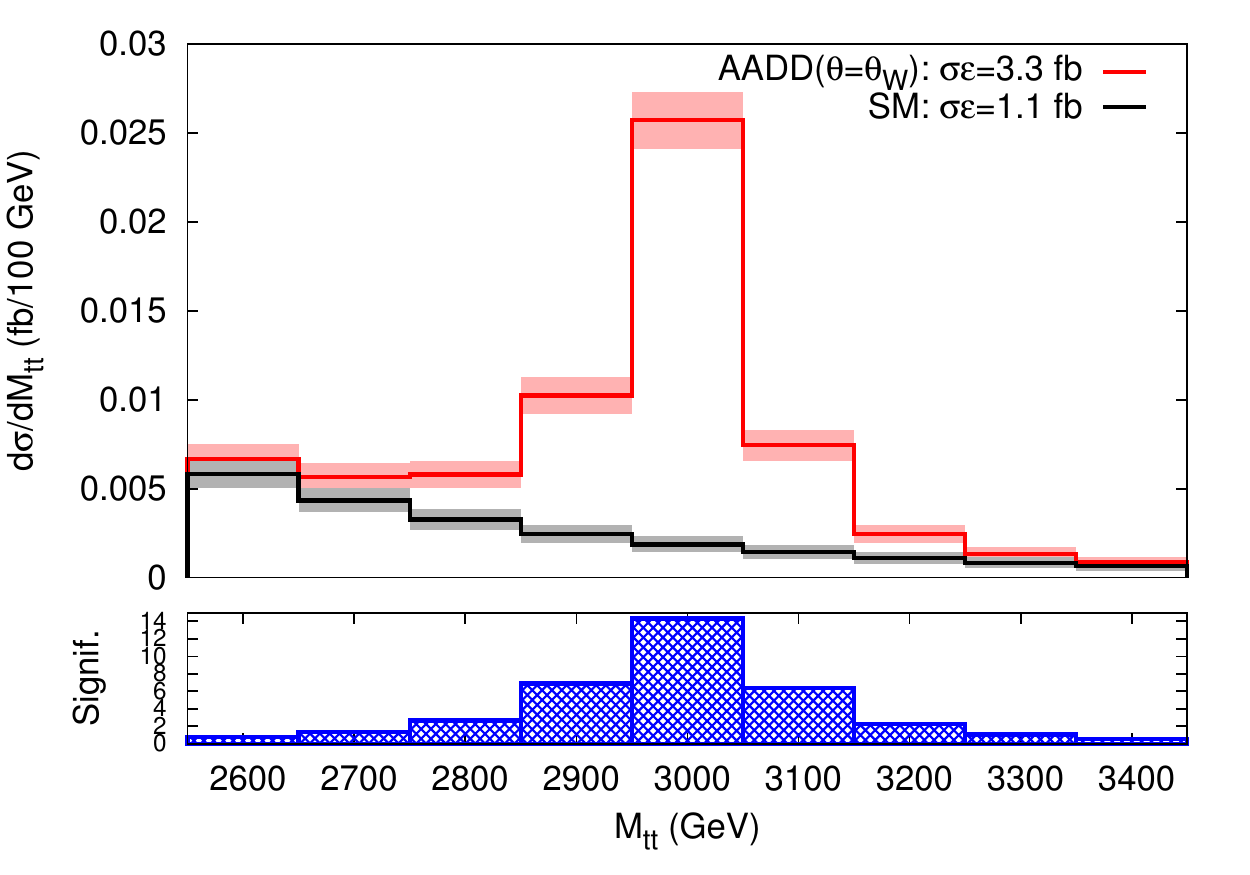}\\
	\includegraphics[width=0.4\linewidth]{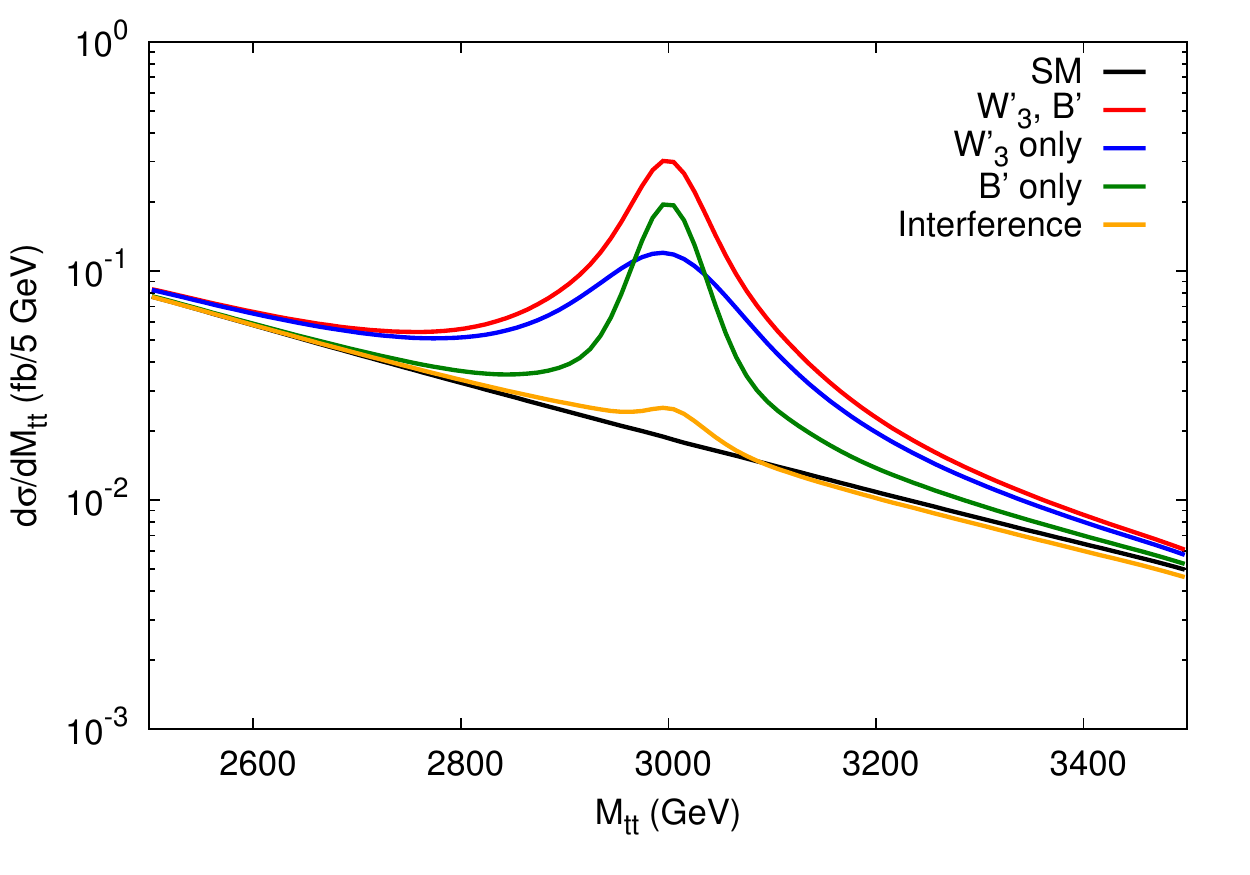}
	\includegraphics[width=0.4\linewidth]{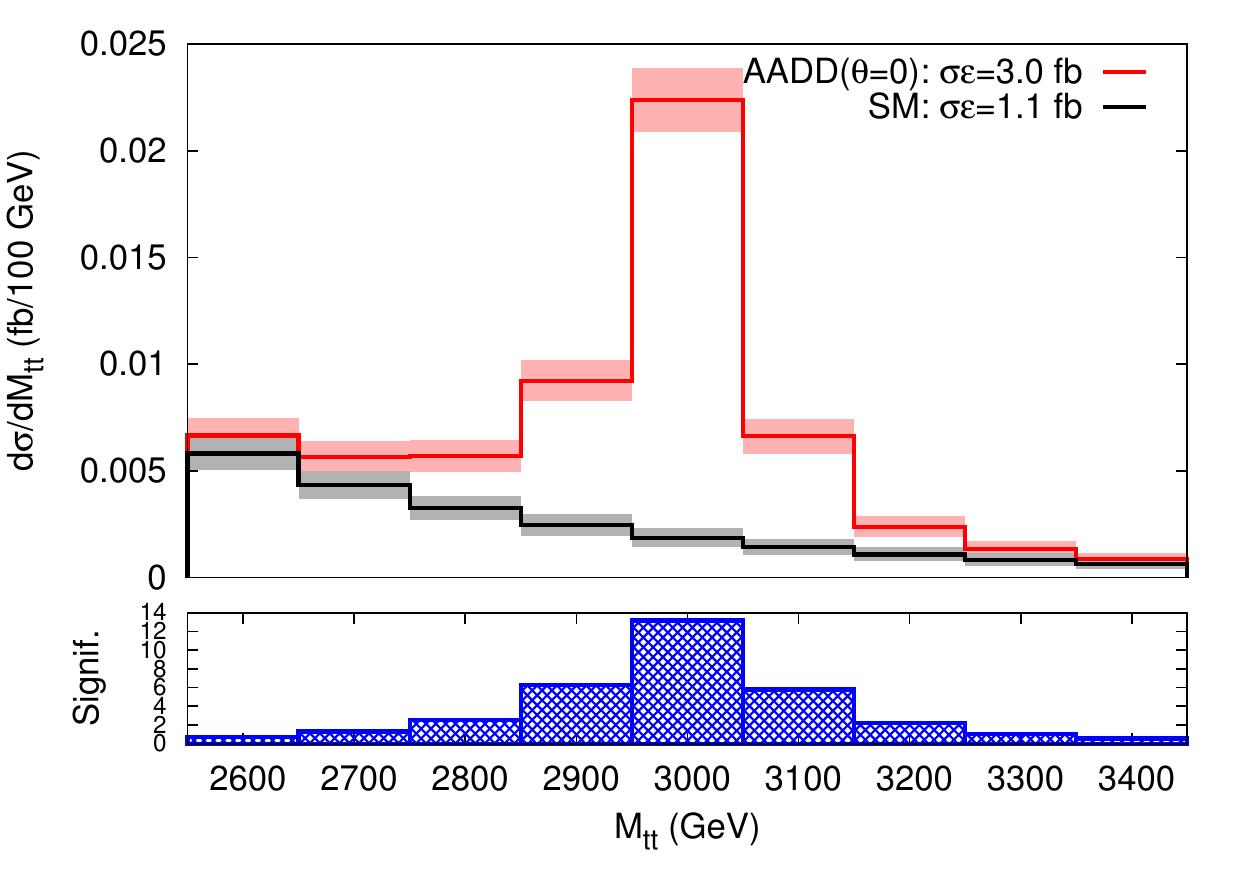}
\caption{The $t\bar t$ invariant mass ($M_{tt}$) distribution of the
cross section for the AADD model with $R^{-1}=$ 3 TeV. The upper two plots show the case where the couplings are $Z$-like and $\gamma$-like while the lower two plots show the case where they are $B$-like and $W_{3}$-like. The left column highlights the contributions from the two resonances and their interference. The right column shows the observables as they would be observed at the LHC at 14 TeV, with 100 fb$^{-1}$ of integrated luminosity, incorporating a 10\% reconstruction efficiency on the $t\bar{t}$ system and statistical uncertainties. The lower subplots on the right hand side measure the bin-by-bin significance of the signal in standard deviations.\label{fig:mtt}}
\end{figure}

We present invariant mass profiles in the standard cross section as well as charge and spin asymmetries for both SM-like `mixed' ($\theta=\theta_{W}$) and the pure `unmixed' ($\theta=0$) case for the LHC at 14 TeV. The relative contributions of the two resonances to the aforementioned observables are decomposed to highlight the fact that, while the invariant mass spectrum views these as a single bump, the asymmetries may allow one to deduce the presence of multiple states. As discussed in Subsect.~\ref{subsec:mass_corrections}, the mixing parameter, $\theta$, should not be physical in the degenerate limit. This appears to be the case for the invariant mass spectra in Figure~\ref{fig:mtt}, where the observable quantity in black reveals the presence of a single resonance, with both contributions and their interference adding coherently to form a Breit-Wigner-like peak. The predictions for both mixed and unmixed cases are rather similar, differing by less than 10\%. The signal ($S$) is, unsurprisingly, very visible above the Background ($B$), as indicated by the large significances, $S/\sqrt{S+B}$, in the right-hand subplots even after folding our estimated 10\% reconstruction efficiency.

\begin{figure}[h!]
	\centering	
	\includegraphics[width=0.4\linewidth]{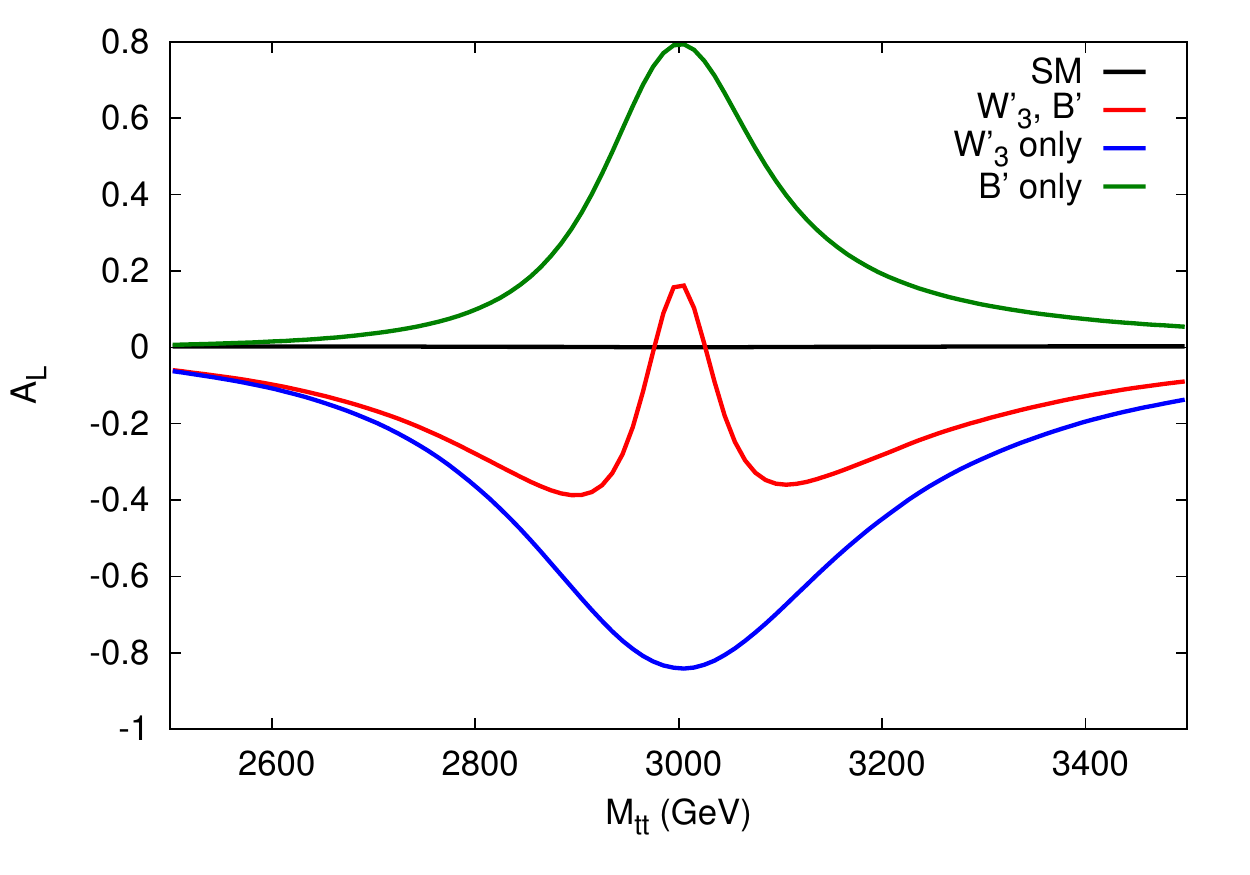}
	\includegraphics[width=0.4\linewidth]{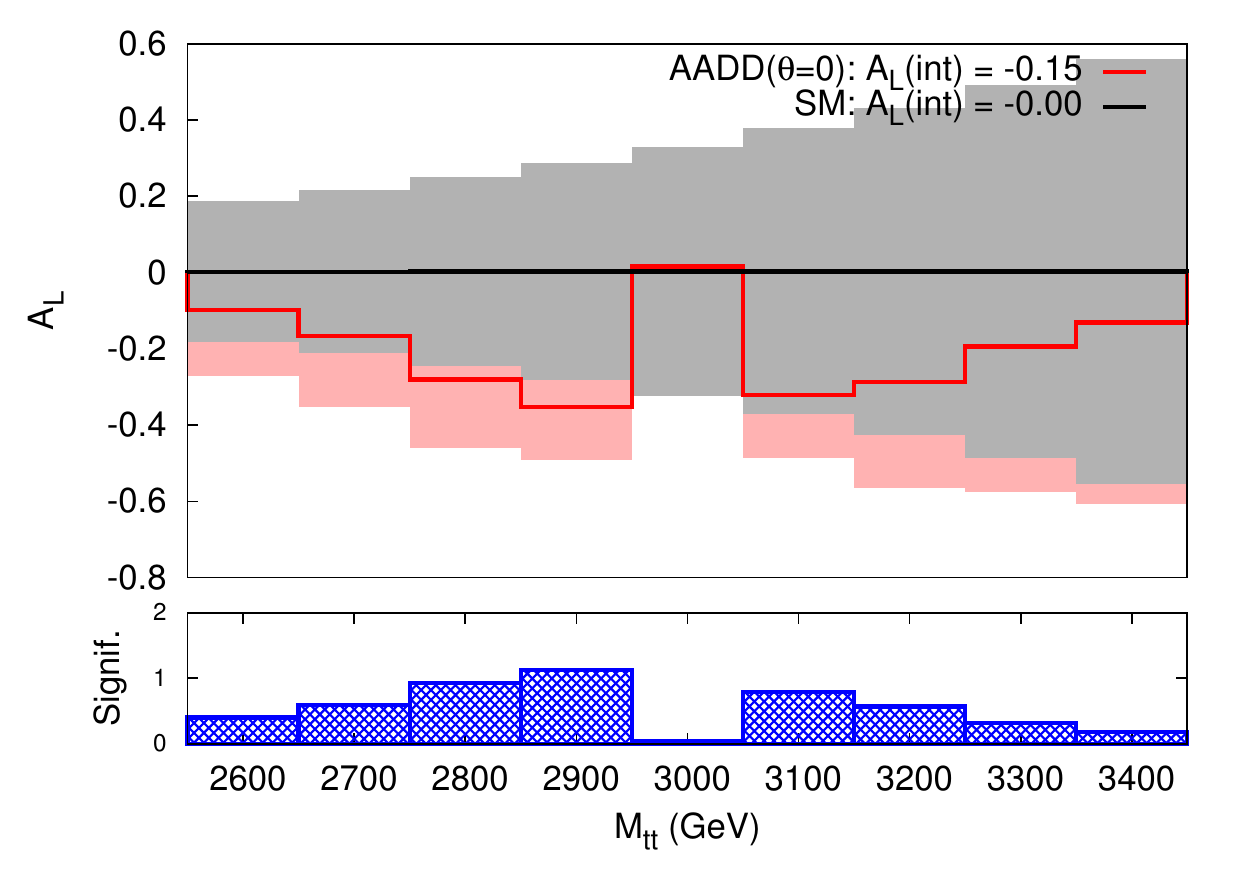}\\
	\includegraphics[width=0.4\linewidth]{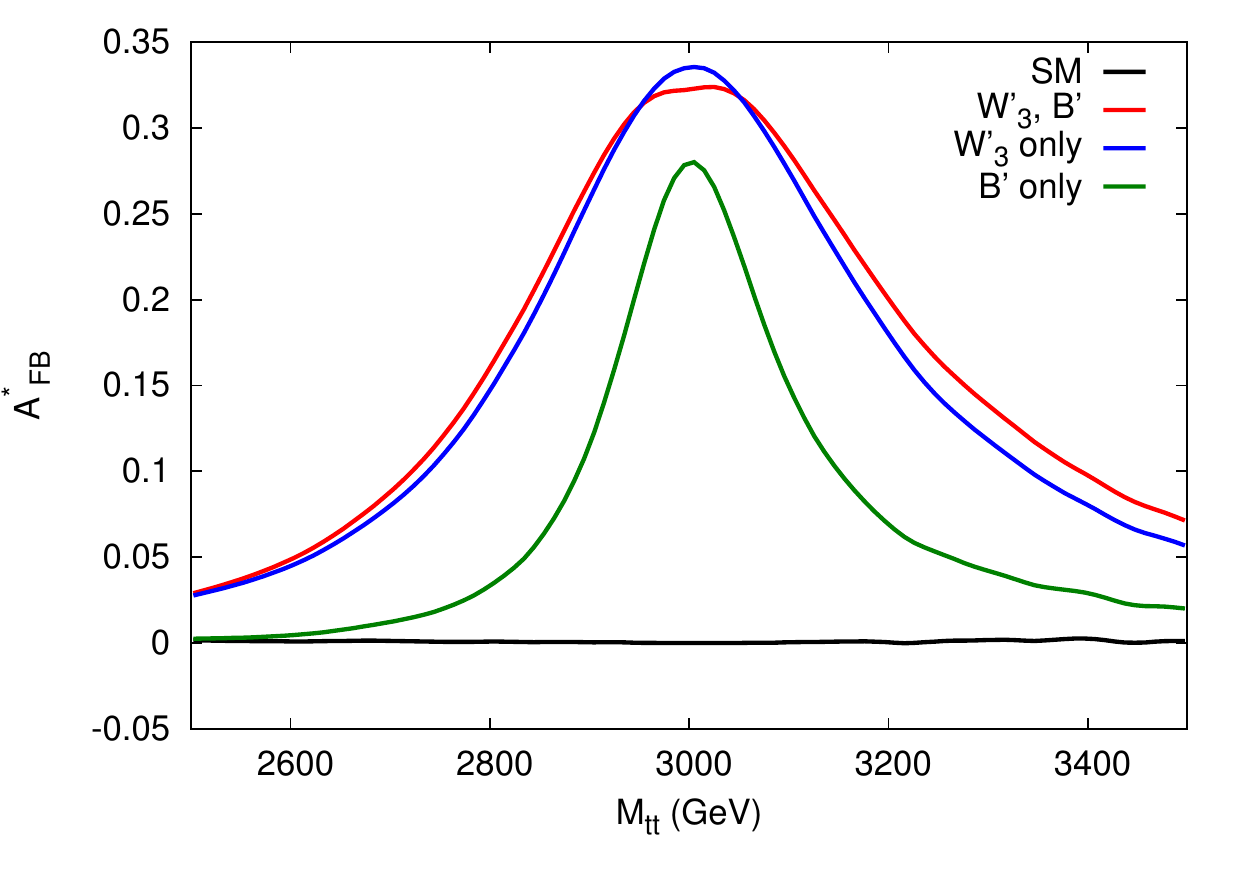}
	\includegraphics[width=0.4\linewidth]{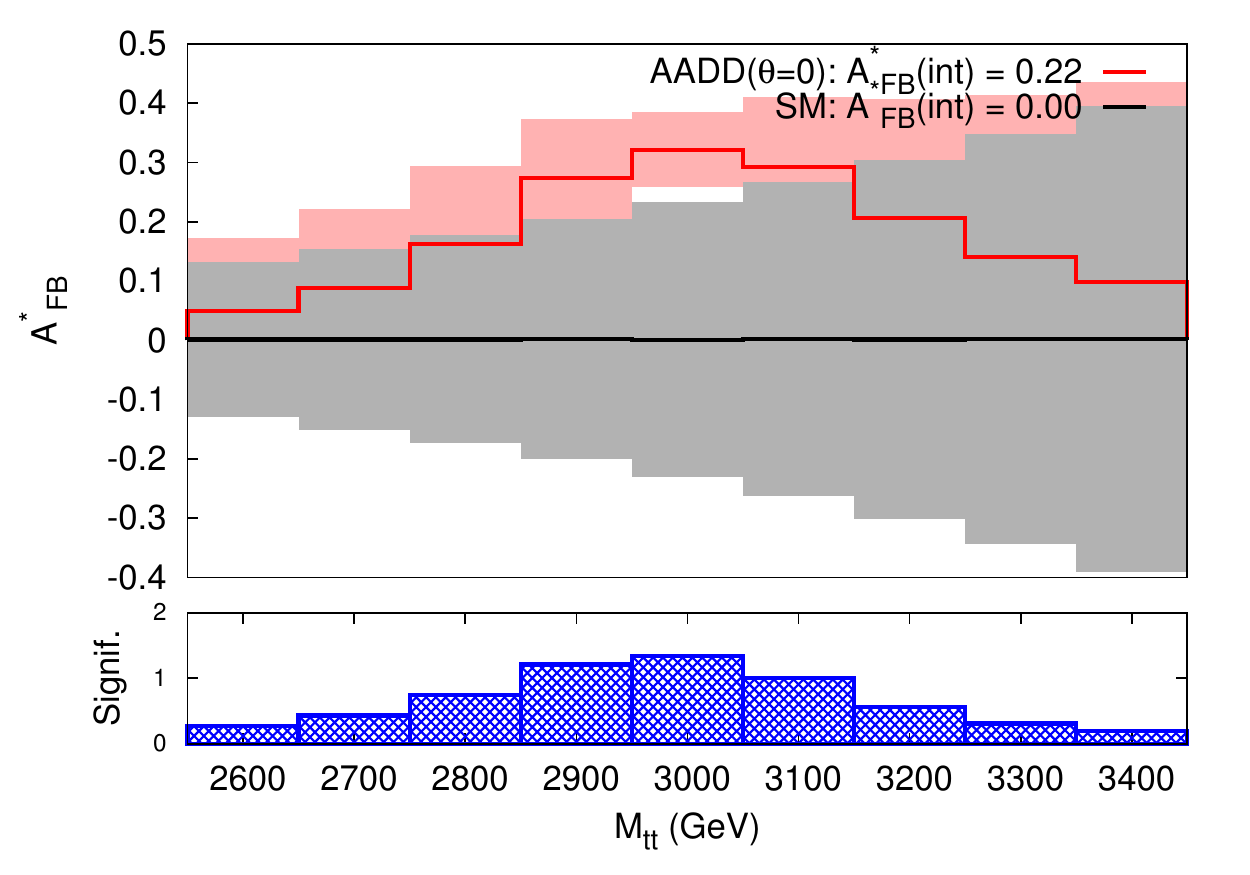}
\caption{The $t\bar t$ invariant mass ($M_{tt}$) distribution of the $A_{L}$ and $A_{FB}^{\ast}$ asymmetries for the AADD model with $R^{-1}=$ 3 TeV where the couplings are $B$-like and $W_{3}$-like ($\theta=0$). The left column shows each of their contributions individually compared to the total (in red). The right column shows the observables as they could be seen at the LHC at 14 TeV, with 100 fb$^{-1}$ of integrated luminosity, incorporating a 10(5)\% reconstruction efficiency on the $t\bar{t}$ system for $A_{FB}^{\ast}(A_{L})$ and statistical uncertainties. The lower subplots on the right hand side measure the bin-by-bin significance of the signal as defined in eq.~\eqref{eqn:signif}.\label{fig:WBasy}}
\end{figure}

\begin{figure}[h!]
	\centering	
	\includegraphics[width=0.4\linewidth]{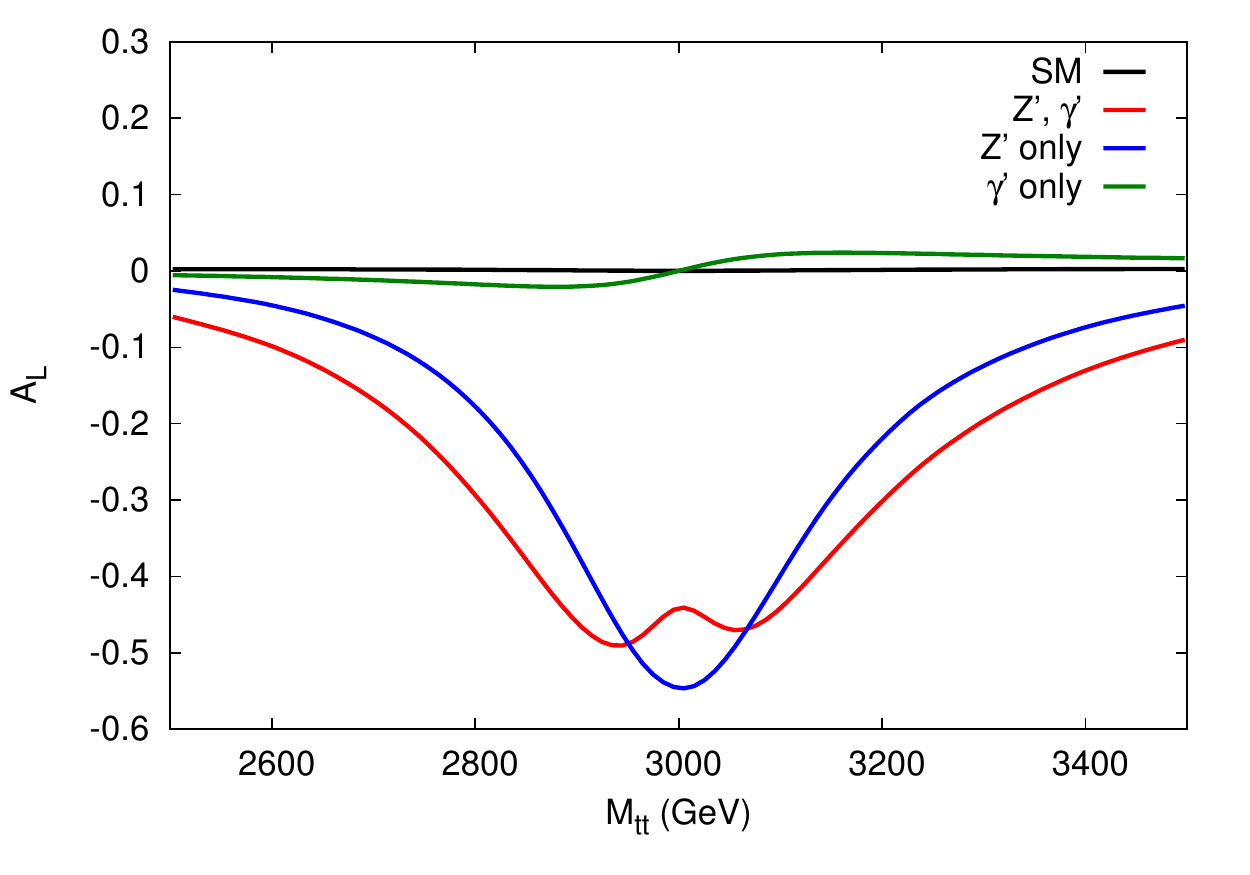}
	\includegraphics[width=0.4\linewidth]{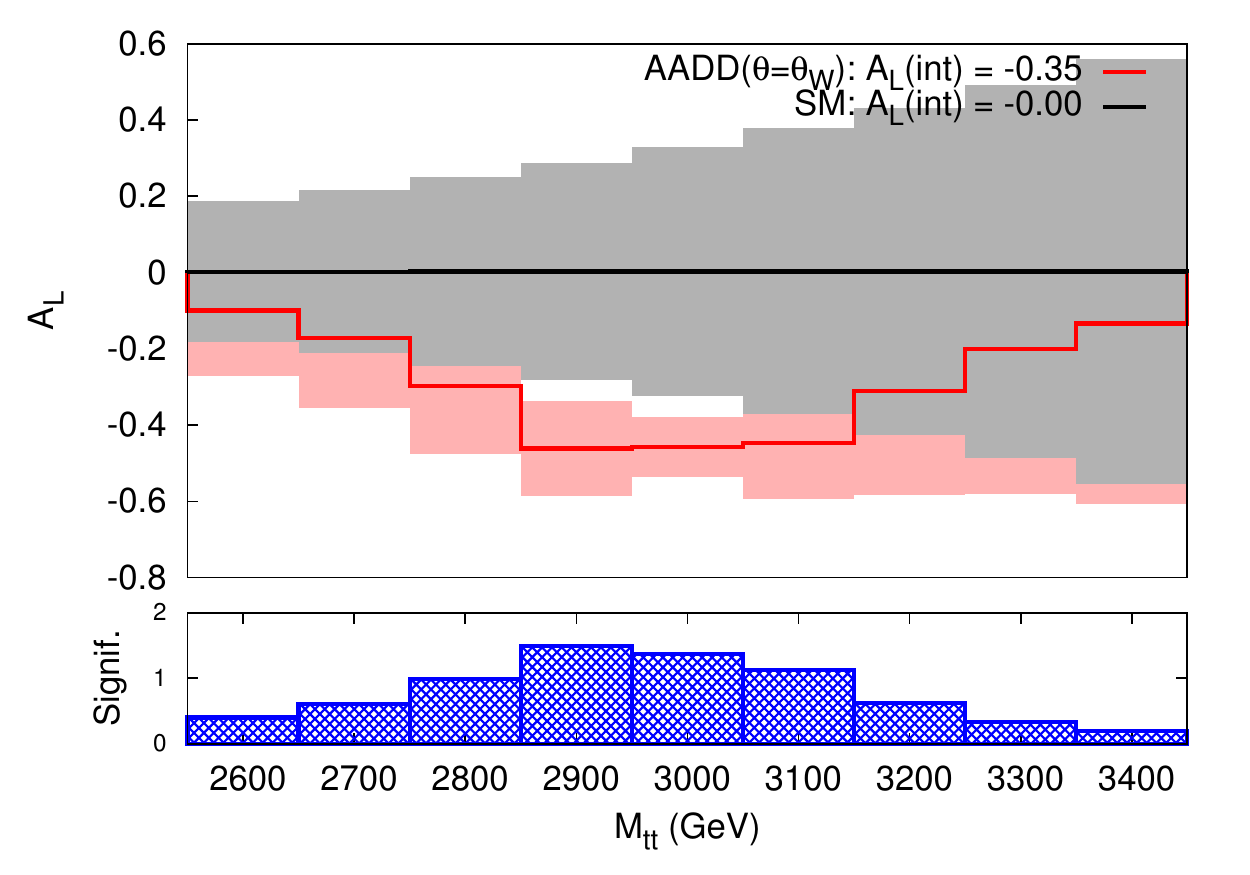}\\
	\includegraphics[width=0.4\linewidth]{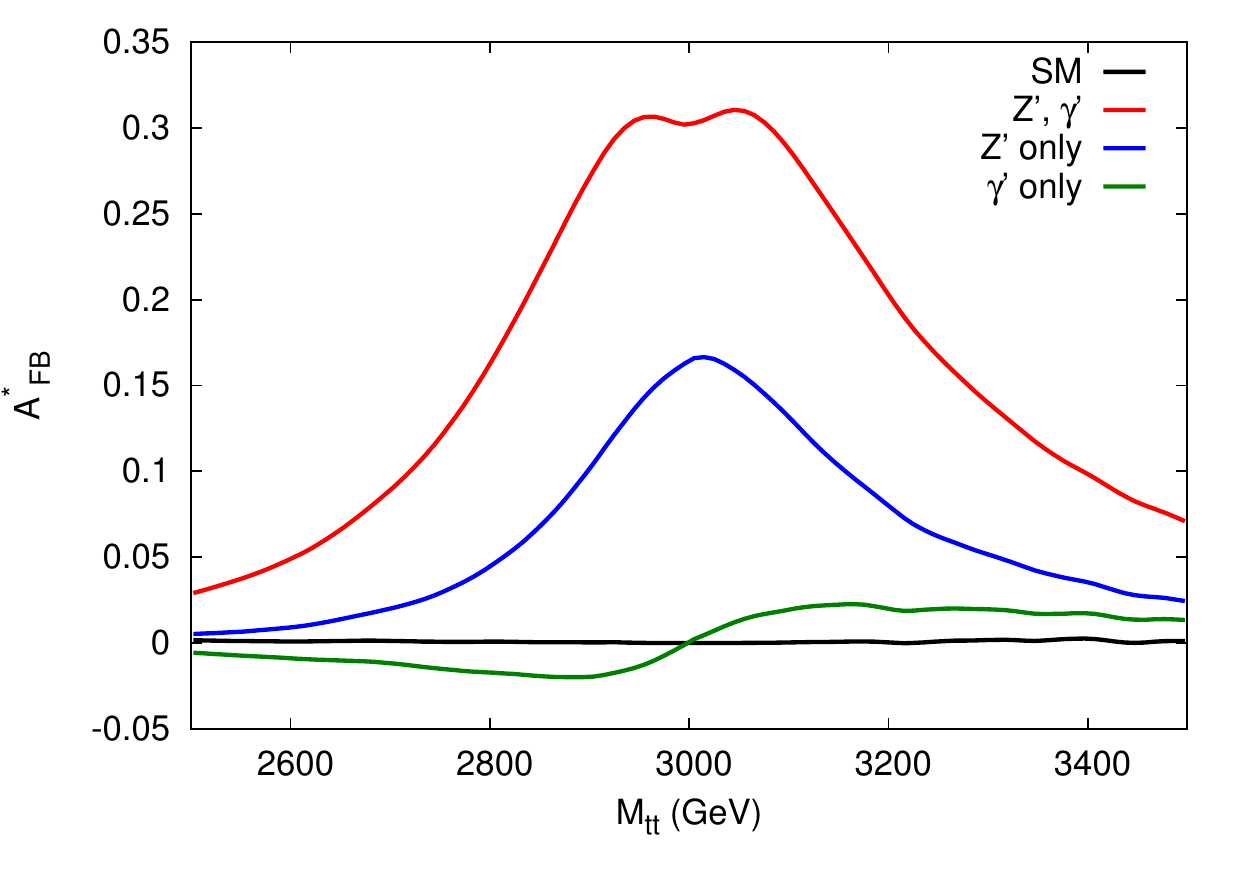}
	\includegraphics[width=0.4\linewidth]{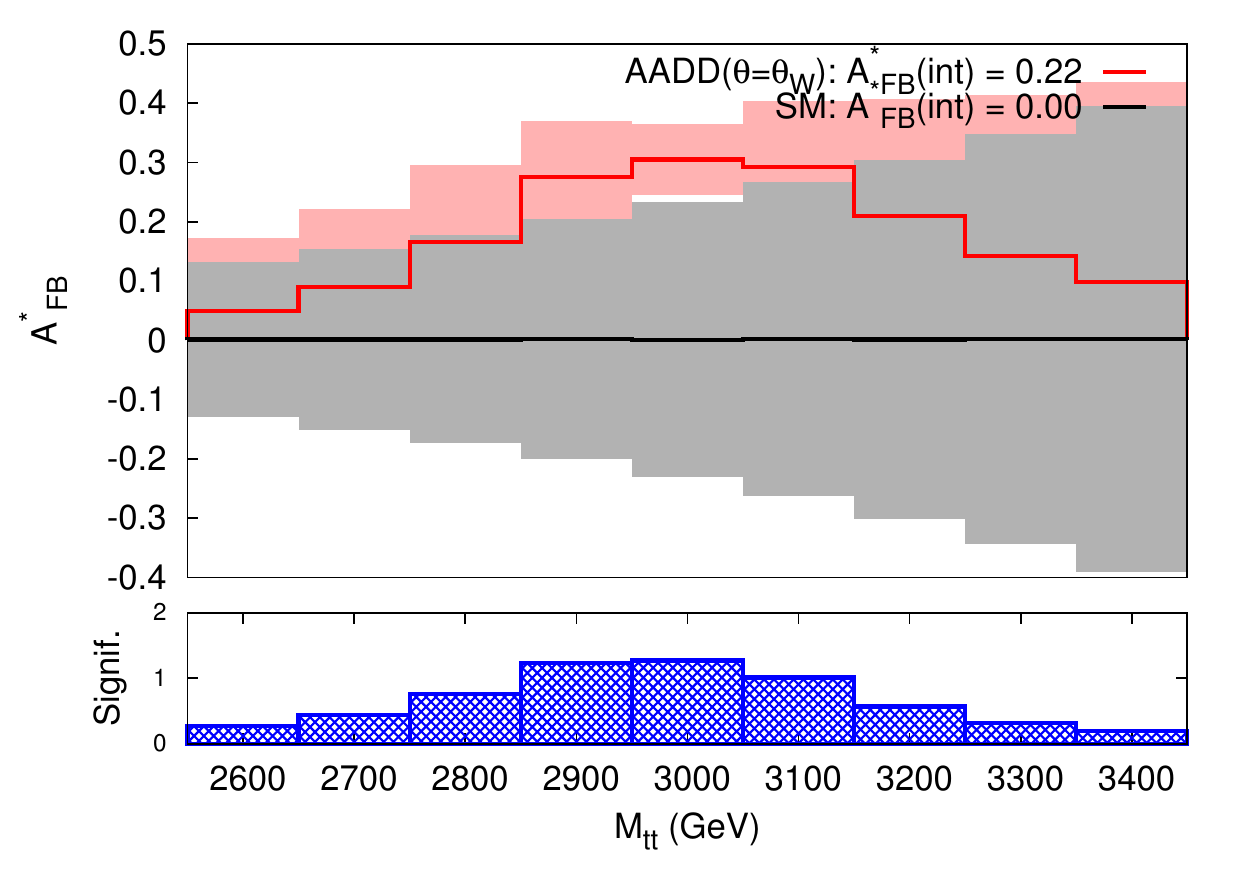}
\caption{Identical plots to Figure~\ref{fig:WBasy} except where the couplings are $Z$-like and $\gamma$-like ($\theta=\theta_{W}$) instead. This corresponds to the unphysical case where off-diagonal matrix elements have not been considered, resulting in differing phenomenology occuring with the variation of an unphysical parameter, $\theta$. \label{fig:ZAasy}}
\end{figure}

\begin{figure}[h!]
	\centering	
	\includegraphics[width=0.32\linewidth]{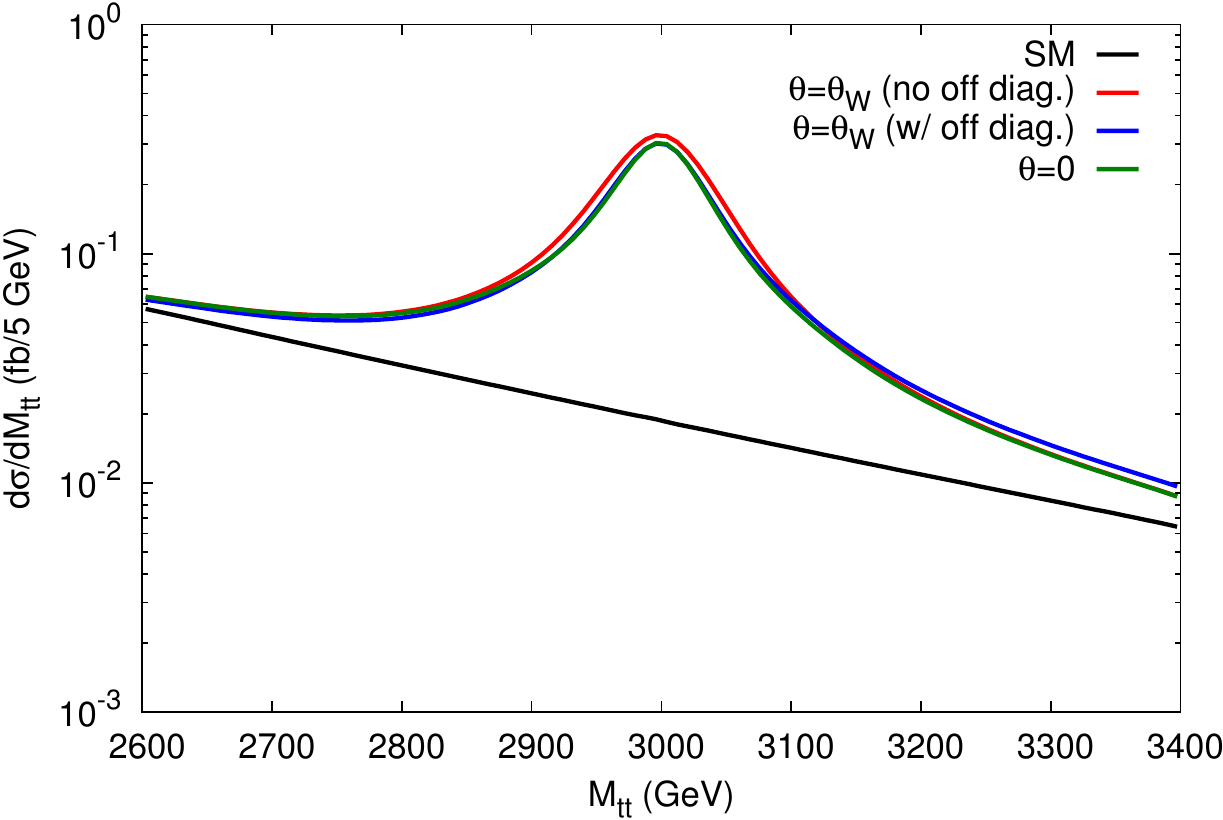}
	\includegraphics[width=0.32\linewidth]{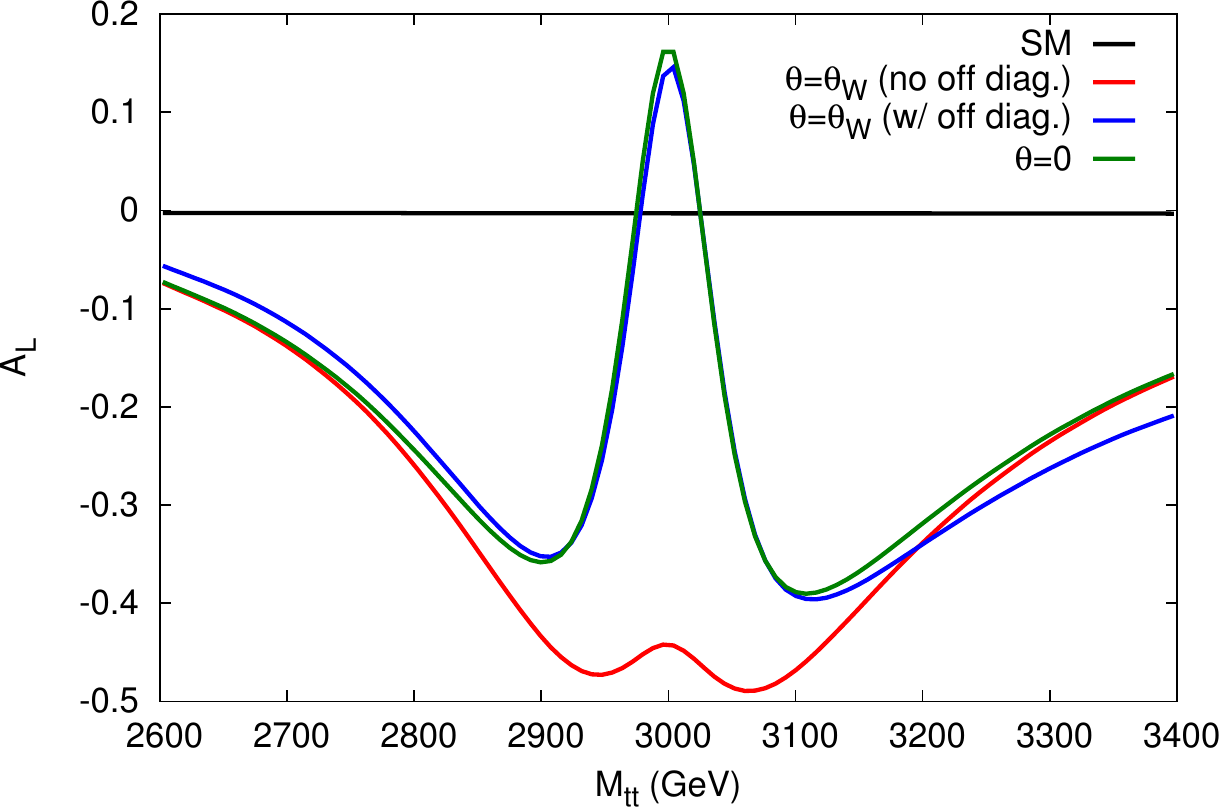}
	\includegraphics[width=0.32\linewidth]{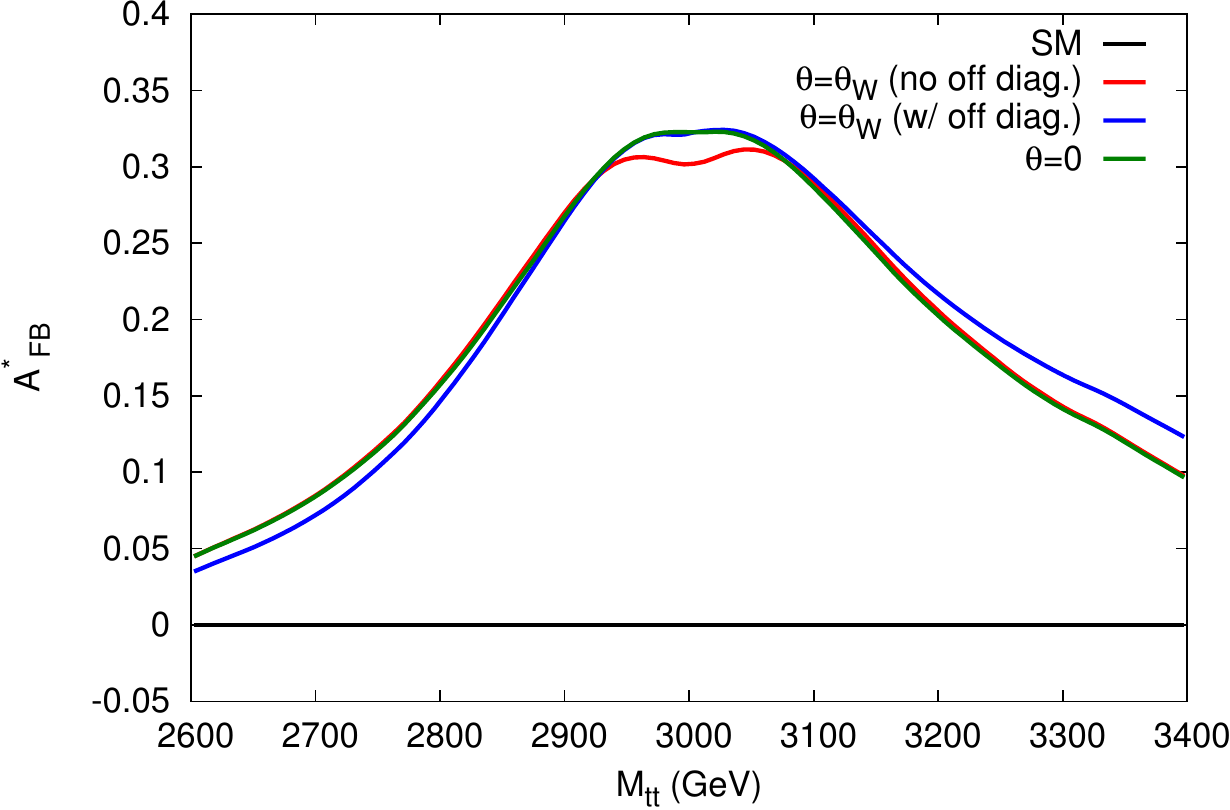}
\caption{Differential distributions in $M_{t\bar{t}}$ for $\sigma$, $A_{L}$ and $A^{\ast}_{FB}$ comparing the mixed AADD with and without off diagonal width contributions to the unmixed case. \label{fig:OD_dists}}
\end{figure}

In contrast, the asymmetries highlight a very different phenomenology. A clear difference can be noted between the prediction for the unmixed and mixed cases in Figures~\ref{fig:WBasy} and~\ref{fig:ZAasy} respectively. This is the unphysical artifact coming from the omission of off-diagonal width contributions discussed in Sect.~\ref{subsec:mass_corrections}. Figure~\ref{fig:OD_dists} shows that the inclusion of these effects makes the prediction for the mixed case consistent with that of the unmixed case, where the off-diagonal terms are zero by construction, restoring the mixing angle to an unphysical parameter. The predictions for the unmixed case and the mixed case with off-diagonal widths agree up to small interference effects away from the peak where the off diagonal terms become small and the latter begins to agree with the mixed case without their inclusion. These deviations are more pronounced in the asymmetries and are likely due to our approximation of only considering off diagonal effects in the denegerate first level KK resonances. We therefore analyse the unmixed scenario as representing the `true' observables in this study.

First, we comment on the physical content of Fig.~\ref{fig:WBasy}. In the upper-left plot, we can see that in the case of $A_{L}$, a characteristic dip appears as a consequence of the two superimposed objects having different widths and couplings. The effects from the wider resonance come in around the edges of the deviation, pushing the value of the observable towards the preferred one for its set of couplings while, near the centre of the distribution, the contribution from the narrower resonance pulls it towards the latter's preferred value. This effect is not as evident in the case of $A^{\ast}_{FB}$, shown in the lower left plot of Fig.~\ref{fig:WBasy}, owing to the dominant contribution to the process coming from the up quark initial state. In the limit where only this state contributes, $A^{\ast}_{FB}(t\bar{t})$ is always positive in such a model with universal fermionic interactions, as can be inferred from Sect.~\ref{subsec:asycoup}. In order to give a complete description of asymmetry effects, in the two left-hand side plots of Fig.~\ref{fig:WBasy} the observables $A_L$ and $A_{FB}^\ast$ are decomposed into contributions from each individual resonance plotted alongside their combination compared to the SM, emphasizing the competition between them. The coupling dependence of such observables allows for this special phenomenology and these observables like to be large since the $W_{3}^{\prime}$ couplings are purely left-handed, maximising the parity asymmetric coefficient in eq.~\eqref{eqn:couplings}.  The right-hand side plots of Fig.~\ref{fig:WBasy} display the two observables, $A_L$ and $A_{FB}^\ast$, with statistical uncertainties at the 14 TeV LHC after 100 fb$^{-1}$ of integrated luminosity folding in a 10(5)\% reconstruction efficiency as mentioned in Sect.~\ref{subsec:reco}. The significances in this case are defined as in eq.~\eqref{eqn:signif} and are lower than those of the invariant mass distribution. Nonetheless, the signal range is rather wide and an integrated value of the observable could provide adequate statistical significance to be observable above the background prediction as we shall show later in Sect.~\ref{subsec:scans}.


Although the `dip' feature of the AADD scenario is visible in the binned $A_{L}$ figures, it is about the only thing that suggests a differing phenomenology from that of a single resonance. Furthermore, the large amount of luminosity required to achieve a more statistically significant differential analysis of asymmetry observables that could confirm the presence of multiple resonances indicates that one may need to rely more on integrated quantities. In the next section we will show that the phenomenology of this model, displaying generic features of quasi-degenerate states, will allow it to be statistically separated from single resonance scenarios using only integrated asymmetries.

\begin{figure}[!h]
\centering
\includegraphics[angle=0,width=0.45\textwidth]{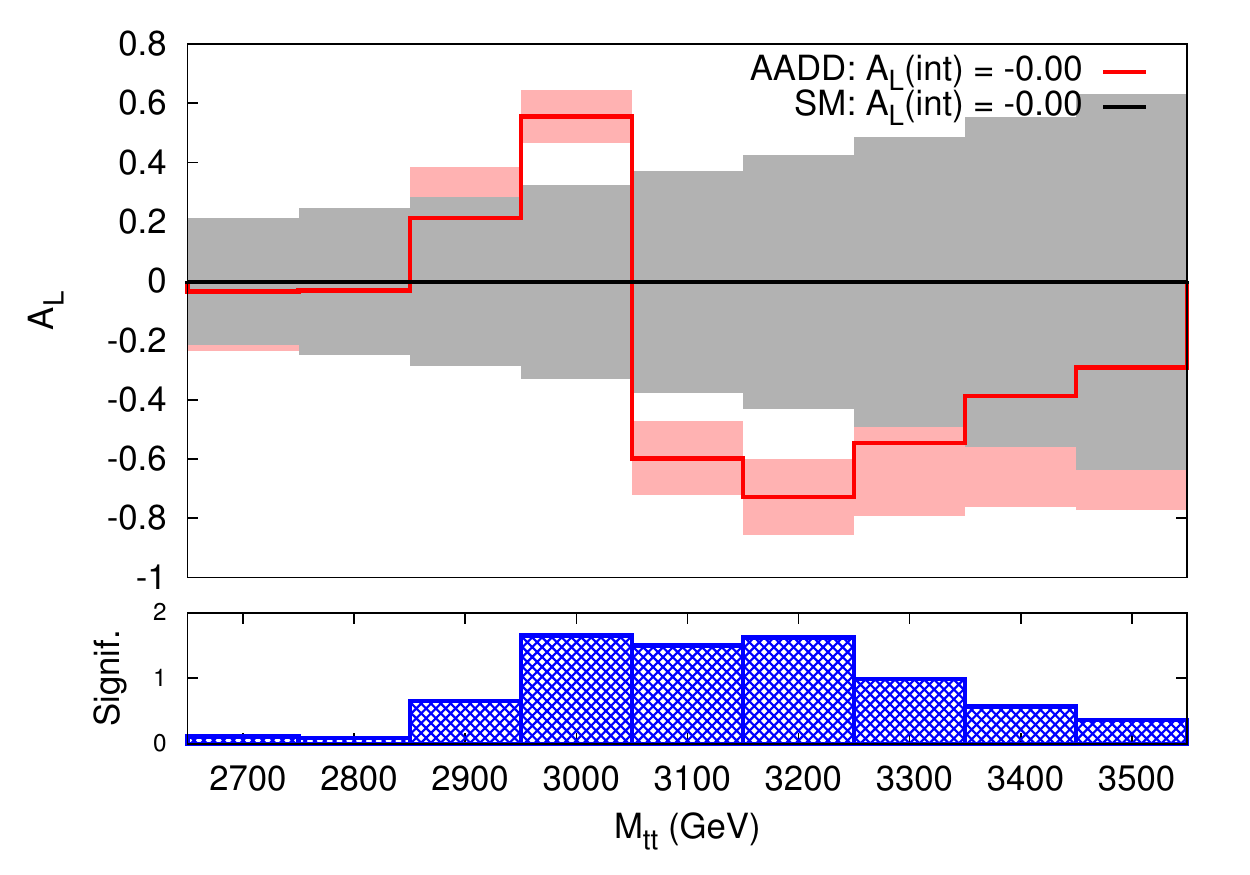}
\includegraphics[angle=0,width=0.45\textwidth]{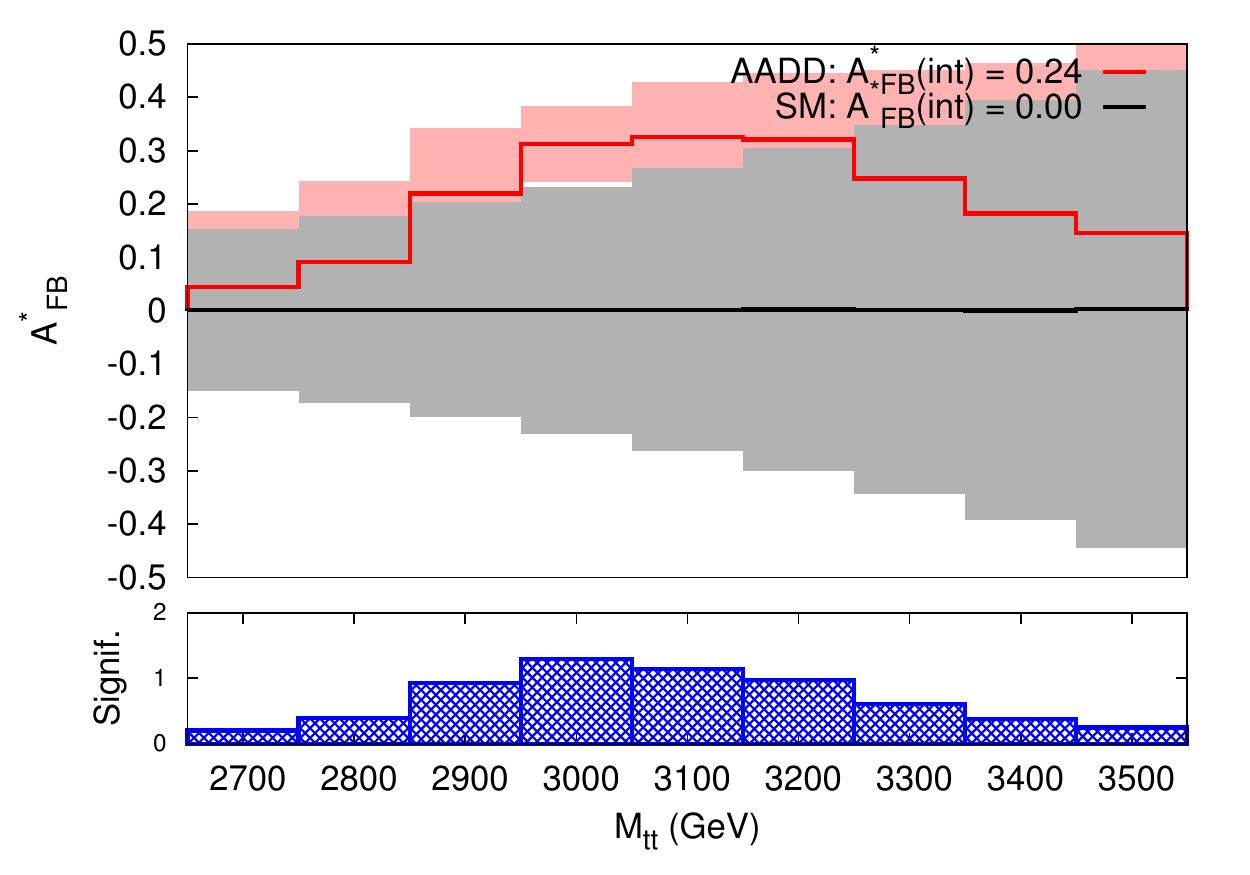}
\includegraphics[angle=0,width=0.45\textwidth]{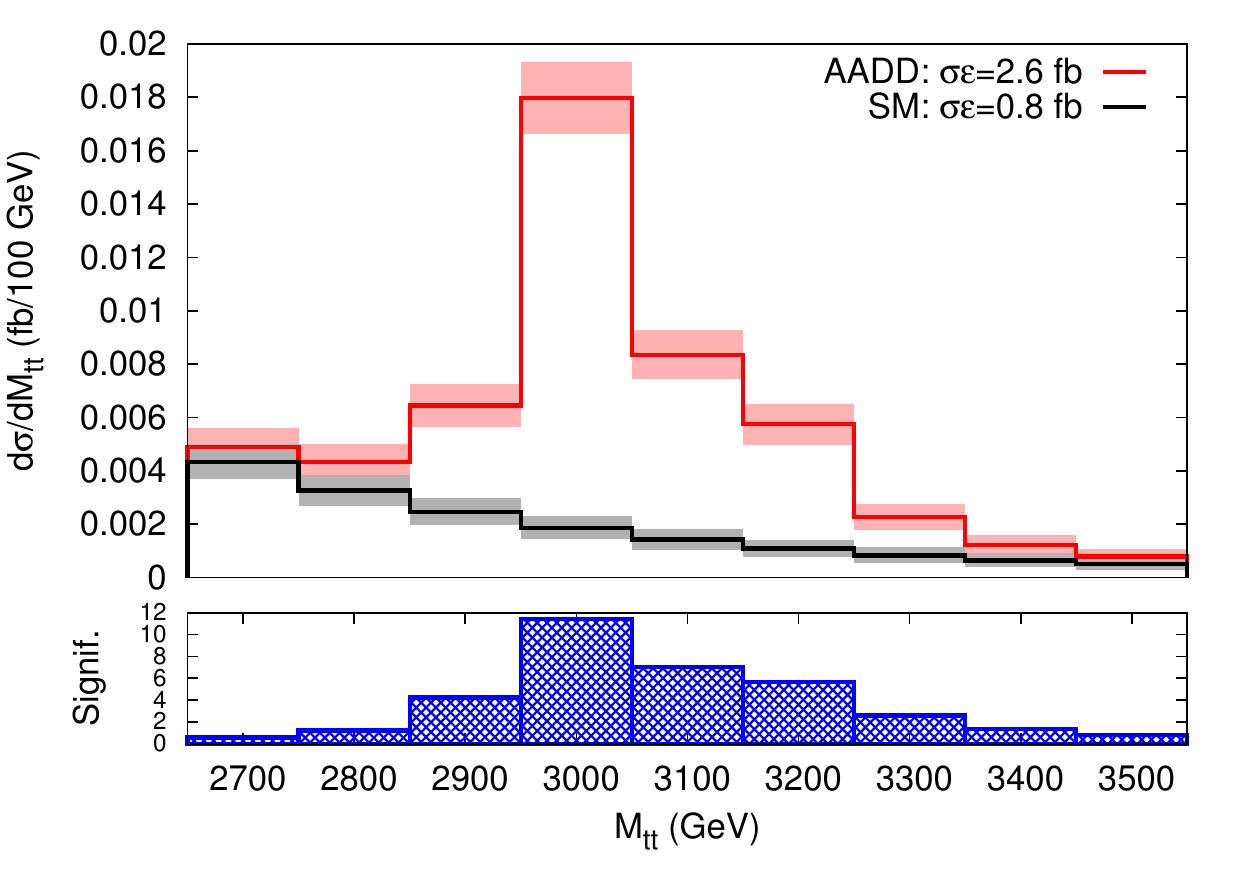} 
\begin{minipage}[b]{0.45\textwidth}
	      \hspace{0.1\linewidth}
      \begin{minipage}[b]{0.9\linewidth}     
 \caption{\label{fig:split_dists}Differential distributions in $M_{t\bar{t}}$ for $\sigma$, $A_{L}$ and $A^{\ast}_{FB}$ for the LHC at 14 TeV, with 100 fb$^{-1}$ of integrated luminosity, incorporating a 10(5)\% reconstruction efficiency on the $t\bar{t}$ system for $A_{FB}^{\ast}(A_{L})$ and statistical uncertainties. The lower subplots measure the bin-by-bin significance of the signal as defined in eq.~\eqref{eqn:signif}.\label{fig:split_distss}}
\vfill
     \end{minipage}
 \end{minipage}
\end{figure}

Before moving to the integrated analysis, we also present the previously shown observables in the split spectrum case ($M_{B^{\prime}}$=2.98 TeV, $M_{W_{3}^{\prime}}$=3.13 TeV), where the radiative mass corrections have been taken into account as described in Sect.~\ref{subsec:mass_corrections}. This drives the mass mixing to zero and brings the model to the edge of the quasi-degenerate regime. Namely, the splitting -- of order 150 GeV -- becomes comparable to the estimated mass resolution and corresponds to about 5\% of $R^{-1}$. We see in Figure~\ref{fig:split_dists} that both the invariant mass distribution and the forward-backward asymmetry still do not resolve two distinct peaks. The spin polarisation asymmetry, $A_{L}$, however, clearly distinguishes between the opposing contributions of the two peaks in an even more striking way than in the degenerate case because the two contributions no longer have to compete at the same invariant mass. Another consequence of this is that the integrated value becomes closer to zero. As we will show in the next section, a single resonance does not generate a forward-backward asymmetry without simultaneously generating a polarisation asymmetry. Thus, the cancellation in the integrated prediction of $A_{L}$ combined with a nonzero $A_{FB}^{\ast}$ will serve as our distinguishing feature.

\subsection{Degeneracy versus a single resonance}\label{subsec:scans}
\begin{figure}[!h]
\centering
\includegraphics[angle=0,width=0.45\textwidth]{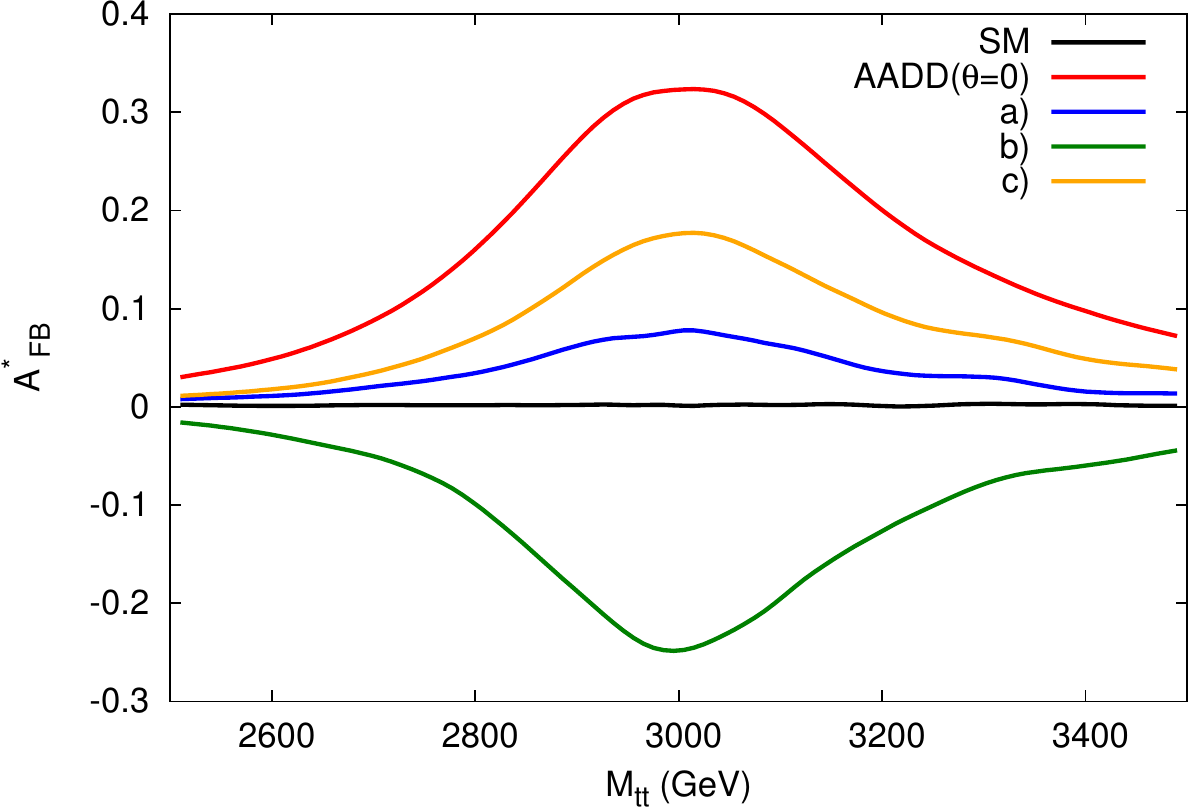}
\includegraphics[angle=0,width=0.45\textwidth]{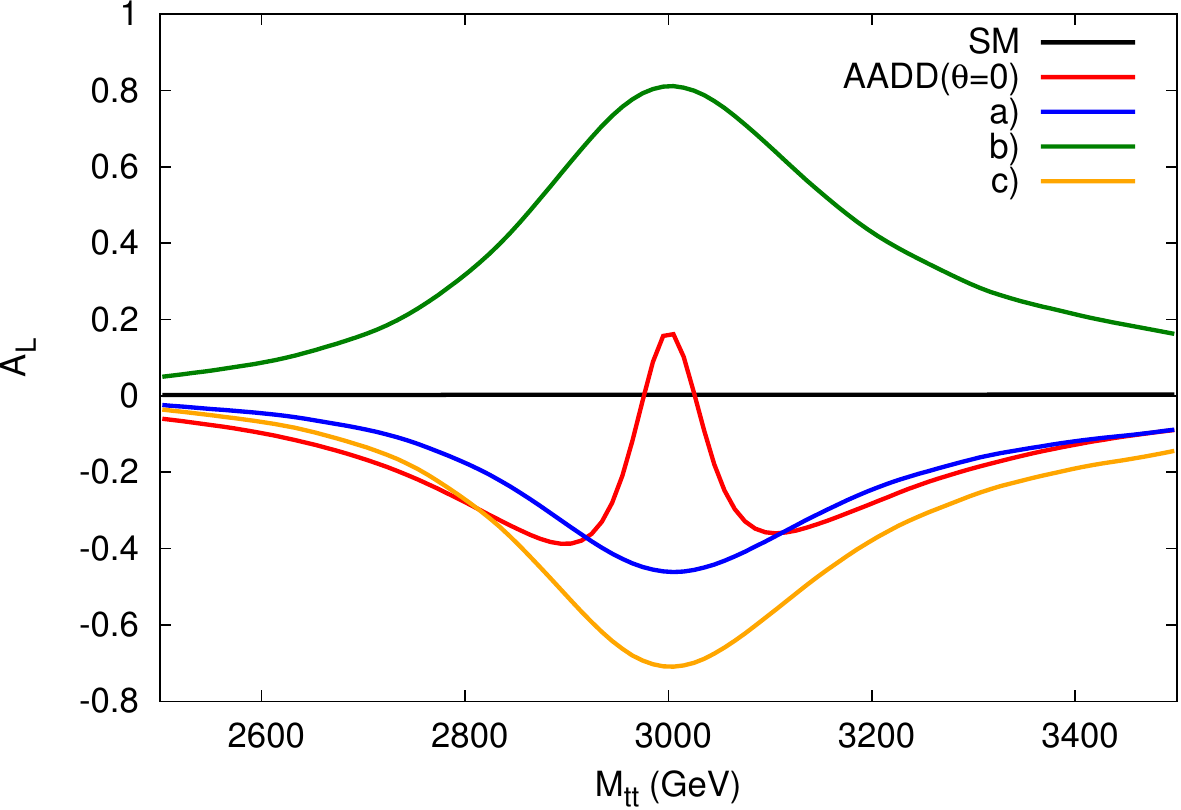}\\
\centering
\hspace{0.4cm}
\begin{minipage}{0.45\linewidth}
\includegraphics[angle=0,width=\linewidth]{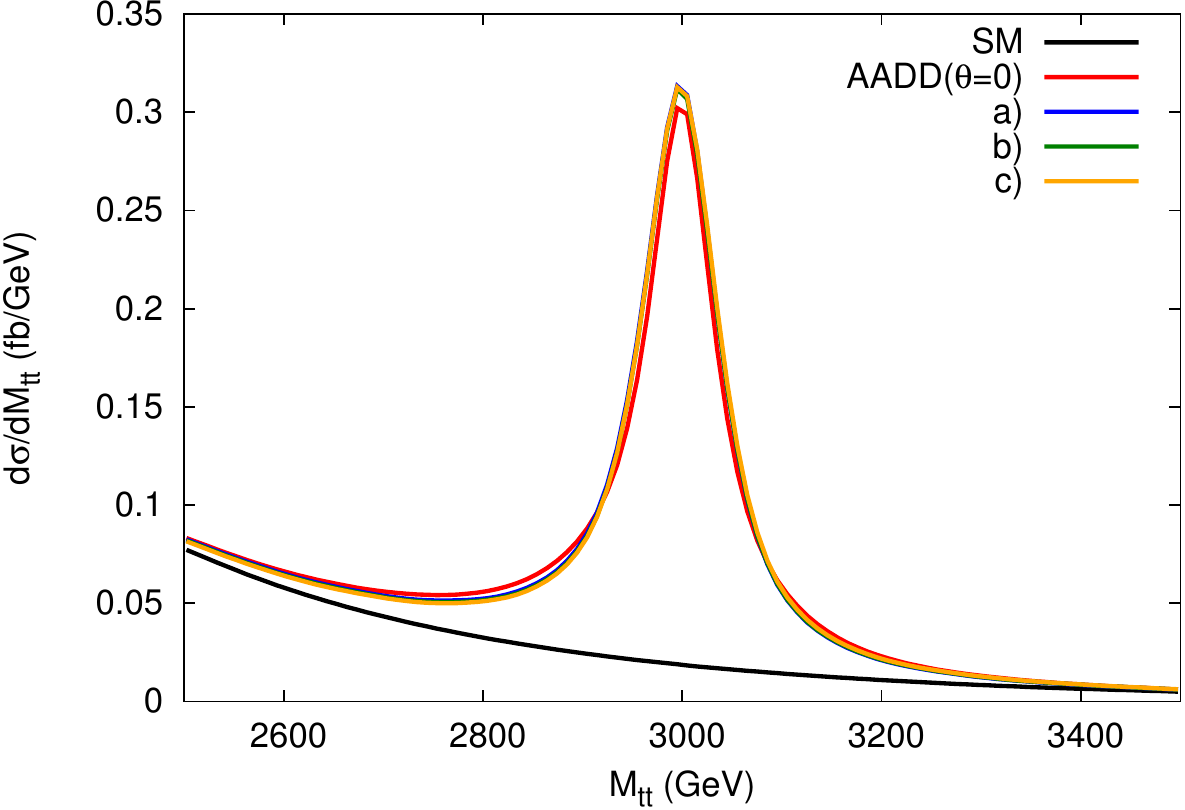} 
\end{minipage}
\hspace{0.5cm}
\begin{minipage}[h]{0.45\textwidth}
        \hspace{0.15cm}
        \small    
         \begin{tabular}[h!]{|c||cccc|}
         \hline
         $\theta=0$&$u_{L}$&$u_{R}$&$d_{L}$&$d_{R}$\tn
         \hline
         \hline
         a)&0.108&0.186&0.846&1.426\tn
         b)&0.242&0.064&0.065&1.401\tn
         c)&0.149&0.405&0.292&0.502\tn
         \hline
         \end{tabular}
         \vfill
 \end{minipage}
 \caption{Differential distributions in $M_{t\bar{t}}$ for $\sigma$, $A_{L}$ and $A^{\ast}_{FB}$ comparing the AADD with three selected
 scan points modelling a single resonance with random couplings generated with its withs fixed to match the cross section
 of each case of AADD. The randomoly chosen couplings are summarised in the lower right table.\label{fig:scan_dists}}
\end{figure}

Having confirmed that the presence of multiple degenerate resonances alters the phenomenology of asymmetry observables, we can explicitly use this to distinguish AADD from models with a single resonance. In order to provide a testbed for this, we created a set of `toy' models of a single resonance designed to be indistinguishable from the degenerate AADD model in a resonance search. This was done by tuning the widths and the couplings and establishing appropriate parameters such that the invariant mass distribution of the points matched those of the AADD. This is shown in Figure~\ref{fig:scan_dists}, which represents a random selection of 3 points fulfilling these conditions. The minimal assumption of universal couplings across fermion generations was made in order to simplify the parameter scan, leaving only the up and down-type chiral couplings $u_{L,R}$ and $d_{L,R}$ as inputs. The other frequent assumption associated with $Z^\prime$s of fixing the charges of each SM representation was ignored, as requiring $u_{L}=d_{L}$ was over-constraining for a toy model, not necessarily meant to represent a physically motivated scenario coming from any particular gauge group extension. The distributions confirm that there are many possible combinations for values of charge and spin asymmetries for seemingly identical resonance cross sections. This is, of course, not surprising following our discussion of the couplings dependences of the various observables in Sect.~\ref{subsec:asycoup} which also implies that the two asymmetry observables are correlated due to their identical dependence on the final state couplings. Again, we note that the observables in AADD remain distinguishable from any of the lettered benchmarks. 

\begin{figure}[t!]
	\centering	
    \includegraphics[width=0.6\linewidth]{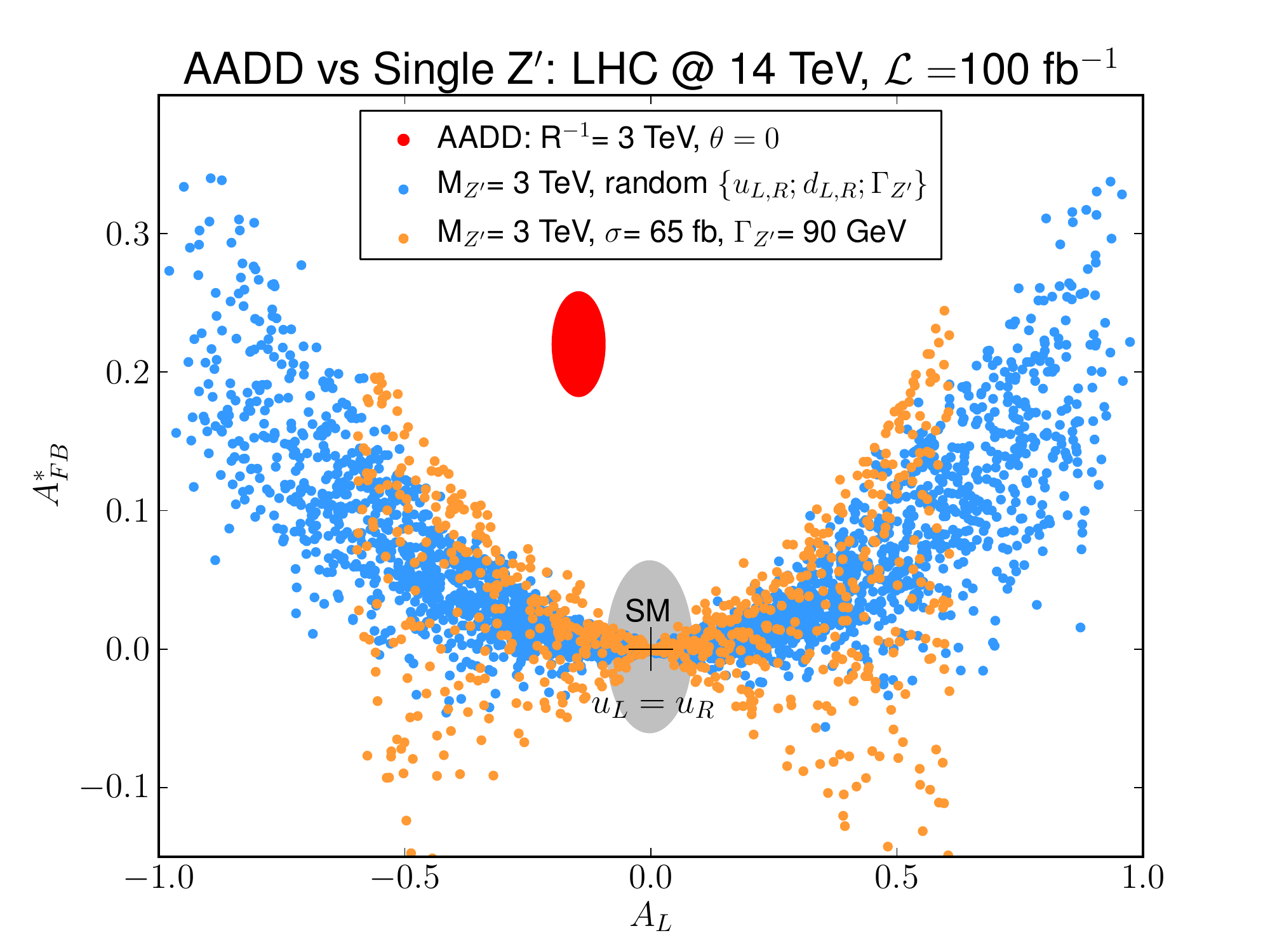}
\caption{Scatter plots showing predicted values of $A_{L}$ and 
$A^{\ast}_{FB}$ for AADD with $R^{-1}=$ 3 TeV 
at the LHC, compared to two sets of points. The first 
represents a scan over random couplings of a single 3 TeV resonance 
with a fixed width constrained to match the AADD invariant mass 
distribution (Figure~\ref{fig:scan_dists}). The second shows a scan 
where the couplings are randomly chosen over the ranges $\{0,1\}$ and 
the resonance width is randomly chosen to be $\leq$ 10\% of the mass. 
The tree-level SM value is shown for reference and ellipses represent 
the 1$\sigma$ statistical uncertainties as defined in 
Sect.~\ref{sec:asymmetries} assuming a 10(5)\% reconstruction efficiency on the $t\bar{t}$ system for $A_{FB}^{\ast}(A_{L})$.\label{fig:scans}}
\end{figure}

With this in mind, we performed a scan over all possible up and down-type couplings allowed while keeping the single resonance cross section (65 fb integrated 500 GeV either side of the resonance) and line-shape (i.e., width) fixed in order to compare and cross-correlate the two asymmetry observables. In addition, we also performed a less constrained parameter scan over any combination of couplings and a random choice of width to see whether the separation power of the asymmetries still holds. The couplings were sampled over an interval $\{0,1\}$ while the widths were chosen to be a random value $\leq$ 10\% of the mass (3 TeV). Both sets of points are shown in Figure~\ref{fig:scans}, where the  AADD case is plotted as an ellipse representing the 1$\sigma$ statistical uncertainties in the asymmetries. The tree-level SM prediction is included for reference, matching the case when the up-type couplings of a single resonance  are purely vector-like ($u_{L}=u_{R}$). The observables plotted are integrated values of the asymmetry over an invariant mass of 500 GeV either side of the resonance mass, for the LHC at 14 TeV and 100 fb$^{-1}$ of integrated luminosity, with statistical uncertainty and reconstruction efficiency estimates consistent with the rest of this study.
 
Firstly, we confirm that the AADD scenario is distinguishable from the SM background in either observable. The profiles of the single resonance scan points show a clear quadratic relationship between the two observables. This can be understood if one assumes that the up quark initial state dominates the production: $A_{L}$ will be proportional to the parity asymmetric coupling combination while $A^{\ast}_{FB}$ will go as the square of this quantity as discussed in Sect.~\ref{subsec:asycoup}. In the case where the invariant mass distribution was constrained to match the AADD rate, the maximum values of $A_{L}$ and $A^{\ast}_{FB}$  are bounded by the maximum absolute value of the couplings. In the unconstrained scan, with the area covered by the points widens slightly due to the larger possible $S/B$, $A_{L}$ becoming unbounded while $A_{FB}$ is limited to be positive and somewhat less than $A_{L}$. This can, again, follows from the coupling dependence of both observables. The parameter scans show that the AADD resonances, in the degenerate limit, can be fully disentangled from any possible single resonance that may produce a similar invariant mass profile in a bump-hunt, within our simplified treatment of reconstruction efficiencies and uncertainties. Therefore this suggests that in the scenario that multiple resonances are observed at the LHC but are masked by a quasi-degeneracy, one may be able to use the asymmetry observables to tell that the signal is coming from more than one resonance. Indeed, any signal appearing as a single peak, with asymmetry values outside of the area spanned by the points in Figure~\ref{fig:scans} will be a smoking gun for degenerate multiple-resonance physics.

\section{Conclusions}
\label{sec:summa}
We have established a realistic example of a model (denoted as AADD) of two quasi-degenerate resonances preferentially decaying to $t\bar t$ final states. Furthermore, the presence of the two new particles cannot be distinguished from a generic single resonance scenario in bump-hunt searches. We have explained that the radiative mass corrections are important and induce splittings that bring the model towards the edge of the quasi-degenerate scenario. However, we have calculated them to be about 5\% of the compactification scale, $R^{-1}$, and maintain that the splittings remain below the $t\bar{t}$ and dijet mass resolutions. In our discussion of radiative mass splittings, quasi degeneracy and subsequent mass mixing, we underlined the importance of a correct treatment of off-diagonal width contributions in this regime. By first considering the degenerate limit as a `worst case scenario' for our purposes, we found that the omission of off diagonal-widths led to potentially misleading artifacts which made the mass mixing angle, $\theta$, appear as a physical parameter even though it should not have.  We used the latest LHC results from dijet and $t\bar{t}$ resonance searches to instruct ourselves on rough limits on the compactification scale from resonance searches at the LHC in order to examine a viable model.

Having expanded on the properties of asymmetry observables in terms of the couplings of said new resonances, we have demonstrated that both charge and spin asymmetries are required to distinguish our scenario from not only any singly resonant signal which mimics the invariant mass distribution of the our model but also any possible observed narrow resonance in $t\bar{t}$ searches. This is owed to the unique features of said asymmetries, that cannot be reproduced in the presence of only one resonant state decaying to $t\bar t$ pairs. In fact, this analysis can serve to probe similar models of multiple quasi-degenerate resonances and a prediction for $A_{L},A^{\ast}_{FB}$ from such a model lying outside the possible values for a single resonance is likely and would signal the presence of multiply resonant physics.

All our results have been obtained at parton level, yet in presence of realistic statistical uncertanties and reconstruction efficiencies,
so they should undergo a certain degree of scrutiny in presence of $t\bar t$ decays, parton shower and hadronisation. However, we expect that the main conclusions of our work will not change substantially.  In addition, the likely boosted nature of the top final state may suggest the need for alternative techniques for measuring top polarisation which do not rely on reconstructing the invariant mass of the top pair. It remains to be seen how the upgraded LHC will be able to deal with spin measurments in boosted tops, but what is clear is that, should they manage to measure the quantity with sufficient accuracy, it would shed much light on the coupling structure and potentially degenerate nature of an observed $Z^\prime$.

\vskip0.5cm
\section*{Acknowledgments}
We would like to thank Ignatios Antoniadis, Matthew Brown and Giacomo Cacciapaglia for 
significant help in understanding the many facets of the model studied here.
SM is supported in part through the NExT Institute. KM is funded by STFC.

\end{document}